\documentclass[12pt]{article}
\usepackage{eurosym}
\usepackage{amssymb}
\usepackage{amsfonts}
\usepackage{amsmath}
\usepackage{geometry}
\usepackage{subcaption}
\usepackage{float}
\usepackage{tikz}
\usepackage{graphicx}
\usepackage[onehalfspacing]{setspace}

\newtheorem{theorem}{Theorem}

\newtheorem{corollary}{Corollary}

\newtheorem{definition}{Definition}

\newtheorem{proposition}{Proposition}

\newenvironment{proof}[1][Proof]{\noindent\textbf{#1.} }{\ \rule{0.5em}{0.5em}}
\begin{document}

\title{\textbf{Competing Mechanisms in Games Played Through Agents: Theory
and Experiment\thanks{%
We thank Bradley Ruffle and seminar participants at McMaster University and
the 58th Annual Meetings of the Canadian Economics Association for their
comments and feedback. Han gratefully acknowledges financial support from
Social Sciences and Humanities Research Council of Canada (\#435-2021-0374).
}}}
\author{Seungjin Han\thanks{%
Department of Economics, McMaster University, 1280 Main Street Hamilton, ON,
Canada L8S 4M4. Phone: (905) 525-9140. Email: hansj@mcmaster.ca} \and Andrew
Leal\thanks{%
Department of Economics, McMaster University, 1280 Main Street Hamilton, ON,
Canada L8S 4M4. Phone: (905) 525-9140. Email: leala@mcmaster.ca}}
\maketitle

\begin{abstract}
This paper proposes Competing Mechanism Games Played Through Agents (CMGPTA), an extension of GPTA (Prat and Rustichini (2003)), where competing principals can offer arbitrary mechanisms that specify a transfer schedule for each agent conditional on all agents' messages. We identify the set of equilibrium allocations using deviator-reporting mechanisms (DRMs) on the path and single transfer schedules off the path. We design a lab experiment
implementing DRMs. A majority of the time, agents tell the truth on the identity of a deviating principal, despite potential gains from (tacit) collusion on false reports. As the game progresses, coordination on both truthful and false reports improves. The average predicted probability of collusion on false reports across groups increased from about 9\% at the beginning of the experiment to just under 20\% by the end. However, group heterogeneity
is significant.

\medskip

\noindent \textit{\noindent Keywords}: competing mechanisms, truthful
communication, collusion, experiment

\noindent \textit{JEL classifications}: C92, D43, D82, D86
\end{abstract}

\section{Introduction\label{sec:intro}}

Economic systems commonly feature many principals and agents interacting in some type of market. Prat and Rustichini (2003) study these interactions in what they call Games Played Through Agents (GPTA), where principals offer each agent a transfer schedule specifying a monetary transfer conditional on the agent's action choice under complete information. An important aspect of multi-principal games is that an agent's private information becomes partley endogenous because they observe contracts offered by all principals. As Prat and Rustichini note\footnote{See footnote 4 in Prat and Rustichini (2003).}, principals may wish to incorporate this market information into the design of their own offers, but what competing principals offer to agents is unobservable by the principal. In this paper, we generalize GPTAs to allow principals to offer and compete with arbitrary mechanisms, which specify a transfer schedule for each agent as a function of all agents' messages.
By communicating with agents, principals can to make their terms of trade responsive to changes in terms of trade offered by the competing principals, despite never observing those terms of trade directly.

In practice, there are contracts that incorporate price-matching clauses, most-favored-customer-clauses (MFCs), or
meeting-competition clauses. For example, a price-matching clause in retail competition allows buyers to report the competing sellers' prices for price match. Similarly, when a seller offers a contract that includes an MFC
clause, a buyer is guaranteed terms of trade at least as good as those given by other sellers by triggering a revision in the contract upon the buyer's report when a competitor offers better terms. A pharmaceutical company may offer an MFC clause to health insurers, promising that they will receive the lowest price compared to competing companies that produce similar drugs. These clauses may soften price competition and lead to higher prices, sustained by this price-adjustment mechanism.

This type of tacit collusion occurs using more complex contracts that allow communication on market information, and is independent of repeated game effects because a deviating principal is punished instantly. Therefore, it is important to study (i) what kind of outcomes one can expect when principals are not restricted in terms of communication contracts permit, and (ii) how agents behave in the presence of contracts that are responsive
to communication with agents on market information (e.g., whether agents strategically misreport).

We call the extended game Competing Mechanism Games Played Through Agents (CMGPTA). As in the literature on competing mechanisms, we assume agents observe the mechanisms offered by all principals. After observing them, each agent simultaneously sends one message to each principal without observing messages that other agents send. Messages then determine transfer schedules. Given transfer schedules, each agent chooses their action.\footnote{%
As in GPTA in Prat and Rustichini (2003), agents observe all transfer schedules offered to all the agents when they choose their actions, but they do not observe them when sending messages.} This communication is intended for principals to make the assignment of their transfer schedules responsive to market information, a feature absent in GPTAs.

We first consider a game where principals offer Deviator-Reporting Mechanisms (DRMs) where each agent uses a binary message to report whether or not there is a principal who deviated from their equilibrium mechanism. If a majority\footnote{Note that we employ the (strict) majority rule for the DRMs to induce agents' truthful reporting. Alternatively, since we consider unilateral deviations for equilibrium in agents' reporting strategies, any majority cutoff strictly greater than half of the agents and strictly less than all agents induces the same set of equilibrium allocations. When there are only two agents, as in our experimental setting, the majority rule is equivalent to a unanimity rule.} of agents report that a principal has deviated from their equilibrium mechanism, the DRM implements transfer schedules to agents, one for each agent, designed to punish the deviating principal. When it is profitable to do so, agents may tacitly collude on their reports and assign themselves a more favorable transfer schedule. For example, if a principal's DRM offers higher monetary transfers for actions when a deviator is reported than when no deviator is reported, agents may collude and report that a competing principal deviated from their equilibrium mechanism, even if they had not. Since principals never observe other principals' offers, such reports cannot be verified. To mitigate this possibility, the DRM implements a transfer schedule offering zero for each action if reports are mixed, encouraging truthful reports. The DRM, therefore, is able to punish not only any deviating principals, but also agents for failing to coordinate their reports.

A common feature in the literature on competing mechanisms, including ours, is that we characterize the set of equilibrium allocations based on agents' truthful reports on information asked by principals.\footnote{Messages in competing mechanisms are intended to convey agents' market information to principals to make their terms of trade responsive to changes in terms of trade chosen by other principals. On the contrary, the message game in Nash implementation by a single principal is designed for the full implementation of an allocation, so that there is a unique equilibrium allocation (See Moore (1992) and Maskin (1985)).} We establish a static folk theorem: the set of PIR-AM allocations - allocations that are individually rational for principals (PIR) subject to agents' optimal action choices given transfer schedules (AM) - is the set of truthful equilibria with DRMs. One convenient feature of a truthful equilibrium with DRMs is that it is sustained if and only if no principal has an incentive to deviate to single transfer schedules, one for each agent. Therefore, there is no need to consider DRMs when a principal deviates.

Many allocations can be supported in a truthful equilibrium with DRMs because it only requires the conditions PIR and AM. We show that even in a 2-principal, 2-agent, 2-action game, any of the four possible action pairs can be an equilibrium action pair. Generally, an efficient equilibrium may occur, but it is also possible to have an equilibrium that jointly maximizes principals' payoffs. For example, sellers (principals) can implicitly collude to charge monopoly prices as long as their competitors do not lower their prices. Once a competing seller deviates to lower their price, the DRM triggers punishment by lowering the non-deviators' prices if a majority of buyers report the deviation. In effect, the DRM induces price undercutting instantaneously, since it is the reports of another principal's deviation that triggers new, lower prices. When deviation-triggered prices are sufficiently low, for example set to the competitive, zero-profit level, other principals cannot deviate profitably, and any vector of optimal actions induced by the original DRM offers are implemented, even at monopoly prices.

As robust equilibrium analysis, we consider a more general game where a principal can send a message to himself in an arbitrarily complex mechanism he offers. As shown in Appendix \ref{sec:CMGPTA_Gamma}, the set of equilibrium allocations in this more general game is indeed the same as the set of PIR-AM allocations. Therefore, there is no loss of generality to focus on truthful equilibria with DRMs. We also demonstrate that equilibria in GPTA survive as equilibria in our extended game.

One limitation of our theoretical result is that, while collusion is possible, our theory does not address when agents will attempt collusion. Since agents each observe all mechanisms offered by all principals, they share the same information as to which principal deviates, if any, and what he offers upon deviation. When it is profitable to do so, agents may collude on their reports, sending a majority of false reports and implementing a preferred allocation for themselves. Furthermore, the complexity of DRMs relative to simple, action-contingent transfer schedules broadens the set of equilibria in the extended game. Therefore, it is worthwhile to design lab experiments to examine how agents communicate with principals and what equilibrium allocations would emerge as a result.

However, it is very difficult to design a lab experiment based on the results from the current literature for two reasons. Firstly, messages in canonical mechanisms are very complex (e.g., an infinite sequence of
real numbers in Epstein and Peters (1999)).\footnote{The complexity of messages stems from the infinite regress problem. As a principal, my terms of trade in a contract depend on a competing principal's terms of trade in their contract. Their terms of trade depends on my terms of trade, so on and so forth. This creates the infinite regress problem in describing market information regarding competitors' contracts, terms of trade, etc.} Secondly, the set of equilibrium allocations is not well defined in terms of model primitives because of the complexity of messages and mechanisms in this canonical setting.

Importantly, our equilibrium allocations are characterized by model primitives in contrast to Yamashita (2010) where they are characterized in terms of complex mechanisms. DRMs, the canonical mechanisms identified in our theory, only require binary messages, so it is very simple by comparison. Furthermore, our theoretical results show that it is without loss of generality to consider DRMs rather than arbitrary mechanisms, and without loss of generality to consider deviations to simple action-contingent transfer schedules. It also offers a clean way to observe false reports, since the principal's offers are observable to the researcher and reports can be categorized as truthful or not ex-post. Thus, the DRM is well-suited for experimental studies on competing mechanism games and makes it possible to design lab experiments to examine agents' communication behavior and resulting equilibrium allocations.

We design a lab experiment implementing DRMs under the simplified 2 $\times$ 2 $\times$ 2 environment to shed light on agents' communication behavior and the resulting equilibrium allocations. In particular, we examine whether agents tacitly collude on their reports and how well they coordinate their reports as play progresses. Our experimental investigation provides a first glimpse of whether the concern on agents' truthful reports is justified and empirical evidence in competing mechanism games.

We study two games; one with a unique efficient outcome and another with efficient outcomes corresponding with individually selfish outcomes. Principals make offers in two transfer schedules for the DRM, observe all the announced DRMs, and decide whether to stay with their DRM or offer a single transfer schedule to each agent  instead.\footnote{While a principal's DRM offer or deviation to single transfer schedule occurs simultaneously the original CMGPTA, we choose such a sequential decision in the experiment to maintain the common knowledge assumption on principals' mechanism on the equilibrium path in the simplest way. This design choice does not alter the set of equilibrium allocations.} As in the theory, principals in the experiment never observe the final offered transfer schedules and instead rely on agents' reports. As a baseline, we run sessions with computerized agents as in Ensthaler et al. (2020), which are programmed to follow the truthful agent reporting assumption. We also run sessions with human agents, where collusion and false reports become possible.

While the theory does not predict what outcome is most likely to be realized, we observe that implemented outcomes are efficient more often than random, and that efficient outcomes were the most frequently implemented outcomes in both computer and human agent treatments. In all treatments, median offers by principals for the least preferred of an agent's actions quickly converge to zero. While not an equilibrium condition in our game, this cost minimization is required for equilibria in a 2 $\times$ 2 $\times$ 2 GPTA, and appears as a feature in the final offers under DRMs observed in the lab. Median offers for the agent's other action were positive throughout the experiments, providing evidence against collusion by principals attempting to extract full rents.

Our findings on agent reports show two sides of the story. A majority of the time, agents do tell the truth, even when faced with incentives to collude on false reports. Of all message pairs, 67.3\% were pairs of truthful  reports. Individually, 29.4\% of agents never falsely report, and another 14.7\% sent false reports less than 10\% of the time. This seems to support the validity of equilibrium characterization of static games based on truthful reporting in competing mechanism games (or partial implementation in a single principal's mechanism design problem).

Coordination on reports improved as the number of mixed report pairs decreases with rounds. This coordination also lead to an increasing in double false reports. Incentives have a statistically significant effect on the probability of coordinated false reports, in line with results from Gneezy (2005) that show people are sensitive to their gains when deciding to lie. We do not observe any significant differences attributable to groups with both female agents, contrasting with findings that all female groups tell less lies than all male or mixed groups (Muehlheusser et al., 2015).

In addition to aggregate results, we look within each group to observe whether agents learn about their counterparty agent. There is significant group heterogeneity in reporting. About half of the groups always sent double true reports except for in at most a couple of rounds, while the other half of groups frequently sent mixed reports. Agents in groups that coordinated on double false reports at least once earned about 8\% less than agents in groups who never sent double false reports, highlighting the difficulty of this coordination problem and the role of the zero transfer schedule in DRMs in dissuading false reporting. Controlling for incentives and group heterogeneity, the average predicted probability of double false reports across groups increases from about 9\% at the beginning of the experiment to just under 20\% by the end. For many groups, the probability rises to nearly 40\% by the end, while for some truthful groups the probability of double false reports remains below 5\% even at the last round. This increase suggests some underlying learning about whether partner agents are willing to falsely report and provides evidence on dynamic lying behavior in group settings. Our paper is one of the first, to our knowledge, to provide empirical evidence on play in competing mechanism games, and hints at the limits of the equilibrium characterization based on truthful reporting in competing mechanisms.

Section \ref{sec:GPTA} summarizes the GPTA proposed by Prat and Rustichini (2003). Section \ref{sec:CMGPTA} extends the GPTA to the CMGPTA and considers the DRM game where principals are restricted to offer DRMs. It provides the equilibrium characterization, and shows that there is no loss of generality to restrict principals to offer single transfer schedules for deviation in the CMGPTA relative to DRMs. Section \ref{sec:design} explains our experiment design and Section \ref{sec:results} presents results. We conclude our paper in Section \ref{sec:conclusion}.

\subsection{Literature Review\label{sec:litrev}}

Tacit collusion through contracting with communication was studied in the literature on contract theory (e.g., Hart and Tirole (1990)) and competing mechanisms (e.g., Attar, et. al (2023), Attar, Campioni, and Piaser (2019), Epstein and Peters (1999), Yamashita (2010), Peters and Troncoso-Valverde (2013)). Our paper belongs to the latter, which studies the set of canonical mechanisms that principals can use without loss of generality in various contracting environments, and provides a static folk theorem given the canonical mechanisms.

Our study extends the experimental literature on Games Played Through Agents (GPTAs), most notably the work of Ensthaler et al. (2020), who provide the first empirical test of the Prat and Rustichini (2003) model. In their design, agents are computerized and deterministic, and principals compete via simple action-contingent transfers. Their results show that efficient outcomes are frequently implemented, consistent with our empirical results. While our study shares their structure, we depart significantly by allowing principals to offer arbitrary mechanisms, which enables richer communication and strategic reporting by agents. This extension allows us to explore questions of truthful reporting, endogenous collusion, and dynamic coordination that are outside the scope of the simpler transfer-only framework in Ensthaler et al. (2020). As a result, while both studies validate the efficiency predictions of GPTA under controlled conditions, our findings reveal the limitations of those predictions when richer strategic environments are introduced.

Our experimental findings also contribute to a substantial body of research on dishonesty in collaborative settings. Prior work has established that team incentives significantly foster dishonest behavior (Conrads et al., 2013; Cohen et al., 2009; Weisel and Shalvi, 2015), with mechanisms like joint responsibility and diffusion of guilt facilitating greater dishonesty in groups. In contrast, Castillo et al. (2022) show that groups may be less dishonest when negative externalities are salient, suggesting contextual factors mediate group dishonesty. A large literature also highlights peer effects in unethical behavior, where individuals adapt their conduct based on observed or anticipated behaviors of group members (Ajzenman, 2021; Kroher and Wolbring, 2015; Kocher et al., 2018; Charroin et al., 2022). These studies emphasize how social learning and peer influence shape dishonest actions, providing important insights into group dynamics.

Our study complements this literature by providing direct evidence on dynamic lying behavior in competing mechanism games, a context where agents' messages influence market outcomes through principal-offered mechanisms. We document how agents learn about other agents' willingness to falsely report over repeated interactions, affecting coordination on dishonest reports. Unlike contexts focusing on static team incentives or exogenous peer effects, our experimental design involves endogenous opportunities for tacit collusion through simple binary messages, enabling us to observe the evolution of group heterogeneity and coordination challenges in real-time. Thus, our results extend existing findings by highlighting the dynamic interplay between incentives, learning, and coordination in fostering dishonest behavior in group settings with strategic reporting opportunities.

\section{Games Played Through Agents\label{sec:GPTA}}

The model is set as follows. There are principals $m\in \mathcal{M}$ and agents $n\in \mathcal{N}$. Let $|\mathcal{M}|=M$, $|\mathcal{N}|=N$, and $M,N\geq 2$. Each agent has available to them a set of actions $S_{n}$. Each principal $m$ offers a simple action-contingent transfer schedule $t_{n}^{m}:S_{n}
\rightarrow
%TCIMACRO{\U{211d} }
%BeginExpansion
\mathbb{R}
%EndExpansion
_{+}$ to each agent $n$. Let $T_{n}^{m}$ be the set of all possible transfer
schedules that principal $m$ can offer to agent $n$. Let $T_{n}\equiv \Pi
_{m\in \mathcal{M}}T_{n}^{m}$ with a typical element $t_{n}=(t_{n}^{1},\dots
,t_{n}^{M})\in T_{n}$ being the set of transfer schedules offered to agent $%
n $ from all principals. Let $T^{m}\equiv \Pi _{n\in \mathcal{N}}\ T_{n}^{m}$
with a typical element $t^{m}=(t_{1}^{m},\dots ,t_{N}^{m})\in T^{m}$ being
the transfer schedules offered by principal $m$ to all agents. Let $T\equiv
\Pi _{m\in \mathcal{M}}\ T^{m}$ with a typical element $t=(t^{1},\dots
,t^{M})\in T$ being the transfer schedules offered by all principals to all
agents.

Principals receive (gross) payoffs $G^{m}(s)$ dependent on the vector of actions $s=(s_{1},\dots
,s_{N})\in S\equiv \Pi _{n\in \mathcal{N}}\ S_{n}$ chosen by the agents, with net payoffs $G^{m}(s)-\sum_{n\in
\mathcal{N}}t_{n}^{m}(s_{n})$. Agents receive (gross) payoffs or costs $F_{n}(s_{n})$ directly from their own action, $s_n$, which may be zero. They also receive the sum of transfers promised to them for choosing action $s_{n}$ for net payoffs $F_{n}(s_{n})+\sum_{m\in \mathcal{M}}t_{n}^{m}(s_{n})$.

Prat and Restuchini's GPTA has the following timing. First, principals simultaneously offer transfer schedules to agents. Each agent observes all the transfer schedules offered to all agents. Given transfer schedules offered to agents, each agent chooses their action.

Equilibria are characterized under pure strategies. Pure strategies for principals in this game are simply transfer schedules $t^{m}\in T^{m}$. Taking these schedules as given, an agent's pure strategy is some action $\sigma _{n}:T\rightarrow S_{n}$.

\begin{definition}
\label{def:G-equilibrium}A pure strategy equilibrium (henceforth an
equilibrium) of a GPTA is a pair $(\hat{t},\hat{\sigma})$, with $\hat{t}=\{%
\hat{t}_{n}^{m}\}_{m\in \mathcal{M},n\in \mathcal{N}}$ and $\hat{\sigma}=\{%
\hat{\sigma}_{n}\}_{n\in \mathcal{N}}$ such that:

\begin{enumerate}
\item for every$\ n\in \mathcal{N},\ $and every$\ t\in T,$%
\begin{equation*}
\hat{\sigma}_{n}(t)\in \ \arg \max_{s_{n}\in S_{n}}\ \left[
F_{n}(s_{n})+\sum_{m\in \mathcal{M}}t_{n}^{m}(s_{n})\right] ;
\end{equation*}

\item for every$\ m\in \mathcal{M},\ $given$\ \left\{ \hat{t}^{j}\right\}
_{j\in \mathcal{M\diagdown }\left\{ m\right\} },$
\begin{equation*}
\hat{t}^{m}\in \ \arg \max_{t^{m}\in T^{m}}\ \left[ G^{m}(\hat{\sigma}(t^{m},%
\hat{t}^{-m}))-\sum_{n\in \mathcal{N}}t_{n}^{m}(\hat{\sigma}_{n}(t^{m},\hat{t%
}^{-m}))\right] .
\end{equation*}
\end{enumerate}
\end{definition}

Let $\mathcal{E}^{G}$ be the set of all equilibria of a GPTA. Fix an
equilibrium $(\hat{t},\hat{\sigma})\in \mathcal{E}^{G}$. Its equilibrium
allocation is denoted by
\begin{equation*}
z^{\left( \hat{t},\hat{\sigma}\right) }\equiv \left[ \left\{ \hat{\sigma}%
_{n}\left( \hat{t}\right) \right\} _{n\in \mathcal{N}}\text{, }\left\{ \hat{t%
}_{n}^{m}\left( \hat{\sigma}_{n}\left( \hat{t}\right) \right) \right\}
_{m\in \mathcal{M}\text{, }n\in \mathcal{N}}\right] \in S\times
%TCIMACRO{\U{211d} }
%BeginExpansion
\mathbb{R}
%EndExpansion
_{+}^{M\times N}.
\end{equation*}
The set of all equilibrium allocations of a GPTA is denoted by
\begin{equation*}
Z^{G}\equiv \left\{ z^{\left( \hat{t},\hat{\sigma}\right) }\in S\times
%TCIMACRO{\U{211d} }
%BeginExpansion
\mathbb{R}
%EndExpansion
_{+}^{M\times N}:(\hat{t},\hat{\sigma})\in \mathcal{E}^{G}\right\} .
\end{equation*}
Bernheim and Whinston (1986) introduce the notion of truthful transfers for
the refinement of an equilibrium in common agency. If $\rvert \mathcal{N} \lvert=1$, a transfer
schedule $t^{m}$ is \emph{truthful relative to }$\hat{s}$ if for every $s\in
S$,
\begin{equation*}
t^{m}\left( s\right) =\max \left( 0,t^{m}\left( \hat{s}\right)
+G^{m}(s)-G^{m}\left( \hat{s}\right) \right) .
\end{equation*}
An equilibrium with $\hat{s}$ as the equilibrium action vector is truthful if every
principal's transfer schedule is truthful relative to $\hat{s}$. If the
agent's action is one-dimensional, a truthful transfer schedule $t^{m}$ is
the upper envelop of the horizontal axis and principal $m$'s indifference
curve going through $\left( t^{m}\left( \hat{s}\right) ,\hat{s}\right) $ in
a two-dimensional space with the vertical axis representing monetary
transfers and the horizontal axis representing the agent's actions. The
equilibrium action with truthful transfer schedules is efficient (Bernheim
and Whinston (1986, Theorem 2)). All truthful equilibria are coalition proof
(Bernheim and Whinston (1986, Theorem 3)). The intuition is that truthful
transfers restrict offers on out-of-equilibrium actions not to be too low
with respect to the principals' payoffs and therefore exhaust all gains from
coalitional deviations.

We cannot extend the notion of truthful transfer schedules to more than one
agent because it induces too many equality restrictions on the transfer
matrix. Prat and Restuchini provides a weaker condition for $N\geq 2$.

\begin{definition}
For principal $m,$ $t^{m}=\left\{ t_{n}^{m}\right\} _{n\in \mathcal{N}}$ is
a profile of weakly truthful schedules relative to $\hat{s}$ if
\begin{equation*}
G^{m}\left( \hat{s}\right) -\sum_{n\in \mathcal{N}}t_{n}^{m}\left( \hat{s}
_{n}\right) \geq G^{m}\left( s\right) -\sum_{n\in \mathcal{N}
}t_{n}^{m}\left( s_{n}\right) \text{, }\forall s\in S\text{.}
\end{equation*}
\end{definition}

Weakly truthful equilibria of GPTA's are always efficient (Prat and Restuchini (2003, Proposition 3)). In particular, if each agent has only two actions and cares solely about monetary payoff $(F_n(s_n) = 0 \ \forall s_n \in S_n, n \in \mathcal{N})$, then all equilibria of GPTA's are always weakly truthful (Prat and Restuchini (2003, Corollary 2)). While action outcomes of GPTAs are efficient in the experiments in Ensthaler et. al (2020), weak truthfulness in principal offers turns out to be violated.

\section{Competing mechanisms\label{sec:CMGPTA}}

We relax the assumption in Prat and Rustichini (2003) that a principal can offer only a single transfer schedule to each agent, and instead permit the use of complex mechanisms for negotiating transfer schedules with agents.

A mechanism can be very complex in terms of the nature of communication that it permits. For instance, principal $j$ can ask agents to report the messages they send to the competing principals, transfer schedules assigned by the competing principals, actions that the agent would choose, etc. The reason is that the principal wants to actively respond to market changes. Although realistic, equilibrium analysis then becomes very challenging.

In subsection 3.1, we consider a simple CMGPTA where each principal can offer a deviator-reporting mechanism (DRM). Transfer schedules for agents depend on their reports on the identity of a principal who offers anything other than their equilibrium DRM. This allows us to establish a static folk theorem that characterizes the set of equilibrium allocations in terms of model primitives.

As shown in Appendix \ref{sec:CMGPTA_Gamma}, the set of equilibrium allocations supportable with DRMs is the same as the set of equilibrium allocations supportable by any mechanism where messages are arbitrarily complex and transfer schedules for agents depend not only their messages but also the principal's message to himself. Therefore, there is no loss of generality to focus on DRMs.

\subsection{CMGPTA with Deviator-Reporting Mechanisms\label{sec:CMGPTA_Lamda}}

We first formally introduce deviator-reporting mechanisms (DRMs) and then present the timing of the game. We let $t_{n}^{\circ }$ be the zero transfer schedule for agent $n$, i.e., $t_{n}^{\circ }(s_{n})=0$ for all $n\in \mathcal{N}$ and all $s_{n}\in S_{n}$. Let $\overline{\mathcal{M}}=\mathcal{M}\mathbb{\cup }\left\{ 0\right\}$ be the set of available messages to agents, which is simply the set of principals plus the zero element. $0\in \overline{\mathcal{M}}$ means that no principal has deviated. Principal $m$'s deviator-reporting contract for agent $n$ is defined by a mapping
$\lambda_{n}^{m}:\overline{\mathcal{M}}^{N}\rightarrow T_{n}^{m}$ such that for all $\left( k_{1}^{m},\ldots ,k_{N}^{m}\right) \in \overline{\mathcal{M}}^{N}$
\begin{multline}
\lambda_{n}^{m}\left( k_{1}^{m},\ldots ,k_{N}^{m}\right) =  \label{DRM} \\
\left\{
\begin{array}{cc}
t_{n}^{m,k} & \text{if }\exists k\neq m\text{ such that }\left\vert \left\{
k_{n}^{m}\in \mathcal{M}:k_{n}^{m}=k,\text{ }n\in \mathcal{N}\right\}
\right\vert >\frac{N}{2}; \\
&  \\
t_{n}^{\circ } &
\begin{array}{c}
\text{if }\exists k,k^{\prime }\neq m\text{ such that (i) }k\neq k^{\prime }
\text{ and} \\
\text{(ii) }\left\vert \left\{ k_{n}^{m}\in \overline{\mathcal{M}}
:k_{n}^{m}=k,\text{ }n\in \mathcal{N}\right\} \right\vert =\left\vert
\left\{ k_{n}^{m}\in \overline{\mathcal{M}}:k_{n}^{m}=k^{\prime },\text{ }
n\in \mathcal{N}\right\} \right\vert =\frac{N}{2}
\end{array}
; \\
&  \\
t_{n}^{m} & \text{otherwise.}
\end{array}
\right.
\end{multline}
Let $k^{m}=\left( k_{1}^{m},\ldots ,k_{N}^{m}\right) \in \overline{\mathcal{M}}^{N}$, $k_{n}=\left( k_{n}^{1},\ldots ,k_{n}^{M}\right) \in \overline{\mathcal{M}}^{M},$ and $k=\left( k^{1},\ldots ,k^{M}\right) \in \overline{\mathcal{M}}^{N\times M}.$ Let $\Lambda _{n}^{m}$ be the set of all deviator-reporting contracts that principal $m$ can offer to agent $n$ and let $\Lambda ^{m}\equiv \Pi _{n\in \mathcal{N}}\ \Lambda _{n}^{m}$ be the set of all possible DRMs available for principal $m$.

A DRM is a profile of deviator-reporting contracts $\lambda ^{m}=\left(\lambda _{1}^{m},\ldots ,\lambda_{N}^{m}\right) \in \Lambda ^{m}$. In a DRM, agents are asked to report the identity of the deviating principal if a competing principal deviates from his DRM that he is supposed to offer in equilibrium. In a DRM $\lambda ^{m}=\left( \lambda _{1}^{m},\ldots ,\lambda _{N}^{m}\right) $ with each $\lambda _{n}^{m}$ in (\ref{DRM}), $t^{m}=\left( t_{1}^{m},\ldots ,t_{N}^{m}\right)$ is a profile of transfer schedules that principal $m$ is supposed to offer to agents in equilibrium. If agents report that no one has deviated (i.e., they all report $0$), $t^{m} $ is offered. If a strict majority of agents report $k\neq m$ as the identity of the deviating principal, principal $m$ offers a profile of transfer schedules $t^{m,k}=(t_{1}^{m,k},\ldots ,t_{N}^{m,k})$ designed to punish principal $k$. If half of the agents report $k$ and the other half report $k^{\prime }$ such that $k\neq k^{\prime },$ at least one of $k$ and $k^{\prime }$ is a false report under unilateral deviations. In this case, all agents are punished in that they get zero transfers regardless of their action choices. In the special case where $N=2$, the strict majority rule is equivalent to a unanimous reports rule, and is particularly useful in inducing truthful reports when there are only two agents.

Given any DRM $\lambda ^{m}$, it is always optimal for each agent to send the true message to principal $m$ when the other agents send the true message, on the equilibrium path and off the path following a principal's unilateral deviation.

The timing of the game relative to $\Lambda \equiv \Pi _{m\in \mathcal{M}}\Lambda ^{m}$ is as follows.

\begin{enumerate}
\item Each principal $m$ announces a DRM $\lambda ^{m}=\left( \lambda
_{1}^{m},\ldots ,\lambda _{N}^{m}\right) \in \Lambda ^{m}$.

\item After observing the profile of DRMs $\lambda =\left( \lambda
^{1},\ldots ,\lambda ^{M}\right) $ announced by principals, each agent $n$
privately sends a message $k_{n}^{m}\in \overline{\mathcal{M}}$ to each
principal $m$.

\item Given the transfer schedules, each agent $n$ chooses action $s_{n}\in
S_{n}$.

\item Payoffs are assigned as in GPTA. Principals receive gross payoffs according to $G^m(s)$ minus the sum of payments they make for the actions. Agents receive the sum of payments owed to them for $s_{n}$ plus some direct
action-specific payoff or loss $F_{n}(s_{n})$.
\end{enumerate}

A strategy for principal $m$ is simply $\left( \lambda _{1}^{m},\ldots
,\lambda _{N}^{m}\right) \in \Lambda ^{m}$. Agent $n$'s communication
strategy is a profile of functions $x_{n}=(x_{n}^{1},\dots ,x_{n}^{M})$
where each $x_{n}^{m}$ is a function from $\Lambda $ into $\overline{%
\mathcal{M}}^{m}$. Therefore, for all $\lambda \in \Lambda $, $x_{n}^{m}(\lambda )\in \overline{\mathcal{M}}$ is the message agent $n$ sends to principal $m$. Let $x=(x_{1},\dots ,x_{N})$ be the profile of all
agents' communication strategies. Given $\lambda \in \Lambda $, let $x_{n}(\lambda )=\left\{ x_{n}^{m}(\lambda )\right\} _{m\in \mathcal{M}}$, $x_{-n}(\lambda )=\left\{ x_{i}(\lambda )\right\} _{i\in \mathcal{N\diagdown }
\left\{ n\right\} }$, $x^{m}\left( \lambda
\right) =\left\{ x_{n}^{m}(\lambda )\right\} _{n\in \mathcal{N}}$, $%
x_{-n}^{m}\left( \lambda \right) =\left\{ x_{i}^{m}(\lambda )\right\} _{i\in
\mathcal{N\diagdown }\left\{ n\right\} },$ $\lambda _{n}\left( x(\lambda
)\right) =\left\{ \lambda _{n}^{m}\left( x^{m}(\lambda )\right) \right\}
_{m\in \mathcal{M}},$ $\lambda ^{m}\left( x^{m}(\lambda )\right) =\left\{
\lambda _{n}^{m}\left( x^{m}(\lambda )\right) \right\} _{n\in \mathcal{N}},$
and $\lambda \left( x(\lambda )\right) =\left\{ \lambda ^{m}\left(
x^{m}(\lambda )\right) \right\} _{m\in \mathcal{M}}.$ Notice that messages
that agents send to principal $m$ depend on a profile of DRMs $\lambda $
offered by all principals. Thus, if some principal were to change his DRM,
it may induce agents to send different messages to the other principals.

Each agent $n$'s action strategy is $\sigma _{n}:\Lambda \times \overline{
\mathcal{M}}^{m}\times T\rightarrow S_{n}$, and $\sigma _{n}(\lambda
,k_{n},t)$ is her chosen action when $\lambda $ is a profile of mechanisms, $
k_{n}$ is a profile of messages she sends, and $t_{n}$ is a profile of
transfer schedules assigned to her. A profile of agents' action strategies
is then $\sigma =\left\{ \sigma _{n}\right\} _{n\in \mathcal{N}}$.

Our solution concept is a pure-strategy subgame perfect Nash equilibrium
(simply an equilibrium) as adopted in Prat and Rustihcini (2003). Given $
\{x,\sigma \}$, let $\sigma \left[ \lambda ,x(\lambda ),\lambda \left(
x(\lambda )\right) \right] =\left\{ \sigma _{n}\left[ \lambda ,x_{n}(\lambda
),\lambda \left( x(\lambda )\right) \right] \right\} _{n\in \mathcal{N}}.$
Then, each principal $m$'s payoff is
\begin{equation*}
V^{m}\left( \lambda ,x,\sigma \right) \equiv G^{m}(\sigma \left[ \lambda
,x(\lambda ),\lambda \left( x(\lambda )\right) \right] )-\sum_{n\in \mathcal{%
N}}\lambda _{n}^{m}(x^{m}(\lambda ))(\sigma _{n}\left[ \lambda
,x_{n}(\lambda ),\lambda _{n}\left( x(\lambda )\right) \right] ).
\end{equation*}

\begin{definition}
\label{def:SPNE}$(\hat{\lambda},\hat{x},\hat{\sigma})$ is a $\Lambda$-
equilibrium if it satisfies the following conditions:

\begin{enumerate}
\item for all $n\in \mathcal{N}$ and all $\left( \lambda ,k_{n},t\right) \in
\Lambda \times \overline{\mathcal{M}}^{M}\times T$
\begin{equation*}
\hat{\sigma}_{n}\left( \lambda ,k_{n},t\right) \in \arg \max_{s_{n}\in
S_{n}}\left\{ F_{n}(s_{n})+\sum_{m\in \mathcal{M}}t_{n}^{m}(s_{n})\right\} ,
\end{equation*}

\item for all $n\in \mathcal{N}$ and $\lambda \in \Lambda $,
\begin{equation*}
\hat{x}_{n}\left( \lambda \right) \in \underset{{k_{n}=\left\{
k_{n}^{m}\right\} _{m\in \mathcal{M}}\in \overline{\mathcal{M}}^{M}}}{\arg \max} \left\{
\begin{array}{c}
F_{n}\left[ \hat{\sigma}_{n}\left( \lambda ,c_{n},\lambda
(k_{n},x_{-n}(\lambda ))\right) \right] + \\
\sum_{m\in \mathcal{M}}\lambda _{n}^{m}(k_{n}^{m},x_{-n}^{m}(\lambda ))\left[
\hat{\sigma}_{n}\left( \lambda ,c_{n},\lambda (k_{n},x_{-n}(\lambda
))\right) \right]
\end{array}
\right\} ,
\end{equation*}

\item for all $m\in \mathcal{M}$ and all $\lambda ^{m}\in \Lambda ^{m}$
\begin{equation*}
V^{m}((\hat{\lambda}^{m},\hat{\lambda}^{-m}),\hat{x},\hat{\sigma})\geq V^{m}((
\lambda^{m},\hat{\lambda}^{-m}),\hat{x},\hat{\sigma}).
\end{equation*}
\end{enumerate}
\end{definition}

Note that $\left( \hat{x},\hat{\sigma}\right)$ satisfying Conditions 1 and 2 in Definition \ref{def:SPNE} constitutes a continuation equilibrium given every possible profile of DRMs in $\Lambda $. An equilibrium $(\hat{\lambda}, \hat{x},\hat{\sigma})$ if all agents report the true identity of a principal who unilaterally deviates.

\begin{definition}
A $\Lambda$-equilibrium $(\hat{\lambda},\hat{x},\hat{\sigma})$ is truthful
if, for all $k\in \mathcal{M}$ and all $n\in \mathcal{N}$,
\begin{equation*}
x_{n}^{k}(\lambda ^{m},\hat{\lambda}^{-m})=\left\{
\begin{array}{cc}
0 & \text{if }m=k\text{ and }\lambda ^{m}=\hat{\lambda}^{m}; \\
m & \text{if }m\neq k\text{ and }\lambda ^{m}\neq \hat{\lambda}^{m}.
\end{array}
\right.
\end{equation*}
\end{definition}

Note that the truthfulness of a $\Lambda $ equilibrium is not defined over
transfer schedules but rather over the agent's reports on which principal
deviates.

A few remarks are in order. Firstly, consider a $\Lambda$-equilibrium $(\hat{\lambda},\hat{x},\hat{\sigma})$ that principals and agents are supposed to play. When we say that principal $m$ is a deviator, it means that principal $m$ deviates from his equilibrium DRM $\hat{\lambda}^{m}$. Secondly, a strategy profile $(\hat{\lambda},\hat{x},\hat{\sigma})$ is defined as a $\Lambda$-equilibrium when no player can gain upon their unilateral deviation, following the conventional definition in non-cooperative games. The DRM constructed in this section is tailored to this approach. Given a profile of DRMs, multiple principals may deviate from their DRMs. Then, agents will play a continuation equilibrium upon multiple principals' deviation where agents do not fully reveal the identities of all deviating principals. This is because the message set in DRMs does not allow the agent to report multiple deviating principals. One may consider the extension of DRMs where the agent is allowed to report the identities of all deviating principals, for example as a vector of binary indicators indexed by principals, where a value of 1 in the $m^{th}$ element of the vector means principal $m$ deviated. Nonetheless, the set of equilibrium allocation would not change because $\Lambda$-equilibrium consider whether a principal has an incentive to unilaterally deviate as in non-cooperative games.

\subsubsection{Equilibrium characterization}

Let $\mathcal{E}^{\Lambda }$ be the set of all truthful $\Lambda $
equilibria. Fix a truthful $\Lambda $ equilibrium $(\hat{\lambda},\hat{x},
\hat{\sigma})\in \mathcal{E}^{\Lambda }.$ The equilibrium allocation is then
\begin{equation*}
z^{(\hat{\lambda},\hat{x},\hat{\sigma})}\equiv \left(
\begin{array}{c}
\hat{\sigma}(\hat{\lambda},\hat{x}_{n}(\hat{\lambda}),\hat{\lambda}(\hat{x}(
\hat{\lambda}))), \\
\left[ \hat{\lambda}_{n}^{m}(\hat{x}^{m}(\hat{\lambda}))[\hat{\sigma}^{m}(
\hat{\lambda},\hat{x}^{m}(\hat{\lambda}),\hat{\lambda}(\hat{x}(\hat{\lambda}
))]\right] _{n\in \mathcal{N}\text{, }m\in \mathcal{M}}
\end{array}
\right) \in S\times \mathbb{R}_{+}^{M\times N}
\end{equation*}
For all $t\in T$, define $\hat{S}(t)$ as
\begin{equation*}
\hat{S}(t)\equiv \left\{ (s_{1},\dots ,s_{N}):s_{n}\in \arg \max_{s_{n}\in
S_{n}}\left[ F_{n}(s_{n})+\sum_{m\in \mathcal{M}}t_{n}^{m}(s_{n})\right] \
\forall \ n\right\} .
\end{equation*}
$\hat{S}(t)$ includes all profiles of optimal actions for agents given $t\in
T$. For each principal $m$, define the minmax value of his payoff relative
to $T$ as follows:
\begin{equation*}
\underline{V}^{m}\equiv \min_{t^{-m}\in T^{-m}}\left[ \max_{t^{m}\in
T^{m}}\left\{ \min_{s(t^{-m},t^{m})\in \hat{S}
(t^{-m},t^{m})}G^{m}(s(t^{-m},t^{m}))-\sum_{n\in \mathcal{N}
}t_{n}^{m}(s_{n}(t^{-m},t^{m}))\right\} \right] ,
\end{equation*}
where $T^{-m}\equiv \left( T^{1},\ldots ,T^{m-1},T^{m+1},\ldots
,T^{M}\right) $. To distinguish transfer \textit{amounts} from transfer
\textit{schedules}, let $d_{n}^{m}$ be a transfer amount from principal $m$
to agent $n$. Let $d^{m}=(d_{1}^{m},\dots ,d_{N}^{m})$ be the vector
transfer amounts to agents from principal $m$ and let $d=(d^{1},\dots
,d^{M})$.

An allocation $(s,d)\in S\times R_{+}^{M\times N}$
satisfies the condition AM (Agent Maximization) if
\begin{equation}
\exists \ t\in T\ \text{s.t.}\ s\in \hat{S}(t)\ \text{and}\
d_{n}^{m}=t_{n}^{m}(s_{n})\ \forall \ (n,m)\in \mathcal{N}\times \mathcal{M}
\text{.}  \label{AM}
\end{equation}
The condition AM says that each agent simultaneously chooses an
action that maximizes her payoff given transfer schedules. An allocation $
(s,d)\in S\times R_{+}^{M\times N}$ satisfies the condition PIR
(Principal Individual Rationality) if
\begin{equation}
G^{m}(s)-\sum_{n\in \mathcal{N}}d_{n}^{m}\geq \underline{V}^{m},\ \forall \
m\in \mathcal{M}  \label{PIR}
\end{equation}
The condition PIR specifies the lower bound of each principal $m$
's payoff $\underline{V}^{m}$. Now we define the set of
PIR-AM allocations:
\begin{equation*}
Z^{\ast }\equiv \left\{
\begin{array}{c}
(s,d)\in S\times \mathbb{R}_{+}^{M\times N}: \\
(s,d)\text{ satisfies PIR and AM}
\end{array}
\right\} .
\end{equation*}
The set of all PIR-AM allocations includes all allocations that are individually rational for principals subject to agents' payoff-maximizing action choices given transfer schedules.

We will show that the set of truthful $\Lambda$-equilibrium allocations is the set of all PIR-AM allocations $Z^{\ast }$ later in Theorem \ref{thm:equilibrium_characterization}. To establish that, it is convenient to use the following technical result.

\begin{proposition}
\label{prop:dev_to_transfer_schedules}$(\hat{\lambda},\hat{x},\hat{\sigma})$
is a truthful $\Lambda$-equilibrium if and only if, for all $m\in \mathcal{M%
}$ and all $t^{m}\in T^{m}$%
\begin{equation*}
V^{m}(\hat{\lambda},\hat{x},\hat{\sigma})\geq V^{m}((t^{m},\hat{\lambda}%
^{-m}),\hat{x},\hat{\sigma})
\end{equation*}
\end{proposition}

\begin{proof}
See Appendix \ref{sec:proof:prop_dev_to_schedules}.
\end{proof}

\bigskip

Proposition \ref{prop:dev_to_transfer_schedules} shows that there is no loss of generality to focus on a principal's unilateral deviation to a profile of transfer schedules instead of all possible DRMs to see if there is a profitable deviation to a DRM given the other principals' DRMs. The intuition is clear. Principal $m$'s deviation to a profile of transfer schedules $t^{m}=(t_{1}^{m},\ldots ,t_{N}^{m})$ is equivalent to a deviation to a DRM that assigns $t^{m}=(t_{1}^{m},\ldots ,t_{N}^{m})$ when agents report $m$as the deviating principal to principal $m$. As a result, $t^{m}=(t_{1}^{m},\ldots ,t_{N}^{m})$ is assigned by principal $m$'s DRM in a
truthful continuation equilibrium. On the other hand, when principal $m$ deviates to a DRM that assigns $t^{m}=(t_{1}^{m},\ldots ,t_{N}^{m})$ when agents report $m$ as the deviating principal to principal $m$, the resulting truthful continuation equilibrium can be reproduced when principal $m$ deviates directly to a profile of transfer schedules $t^{m}=(t_{1}^{m},\ldots ,t_{N}^{m})$.

Using Proposition \ref{prop:dev_to_transfer_schedules}, we can establish the following folk theorem like result.

\begin{theorem}[Static Folk Theorem]
\label{thm:equilibrium_characterization} The set of PIR-AM
allocations, $Z^{\ast }$ is the set of truthful $\Lambda$-equilibrium allocations.
\end{theorem}

\begin{proof}
See Appendix \ref{sec:pf:thm:eq_charzn}.
\end{proof}

\bigskip

Theorem \ref{thm:equilibrium_characterization} provides a folk theorem in a static game in the sense that any allocation that is PIR-AM can be supported as a truthful $\Lambda$-equilibrium allocation. The intuition is similar to folk theorems in repeated games. To see this, let us pick any PIR-AM allocation $(s,d)$. Then, there exists a profile of transfers $t\in T$ such that the agent's payoff-maximizing action choices given $t$ leads to $(s,d)$ which induces each principal $m$'s payoff to be no less than $\underline{V}^{m}$. Then, each principal $k$ can offer a DRM where transfer schedules $t^{k}\in T^{k}$ are assigned when a strict majority of agents report no deviation by any other principal, but transfer schedules $\tilde{t}^{m,k}\in T^{k}$ are assigned when a strict majority of agents report $m\neq k$ as the identity of a deviating principal. Therefore, $\left\{ \tilde{t}^{m,k}\right\} _{k\in \mathcal{M} \backslash \{m\}}$ is the profile of transfer schedules that punish principal $m$'s deviation and it satisfies
\begin{multline}
\left\{ \tilde{t}^{m,k}\right\} _{k\in \mathcal{M}\backslash \{m\}}\in \\
\arg \min_{t^{-m}\in T^{-m}}\left[ \max_{t^{m}\in T^{m}}\left\{
\min_{s(t^{-m},t^{m})\in \hat{S}(t^{-m},t^{m})}G^{m}(s(t^{-m},t^{m}))-%
\sum_{n\in \mathcal{N}}t_{n}^{m}(s_{n}(t^{-m},t^{m}))\right\} \right] .
\label{punishment}
\end{multline}
As players minmax a deviating player for punishment in the repeated game, principals also minmax a deviating principal in our static game, given that agents also participate in the punishment (the inner most minimization in (\ref{punishment})).

Without loss of generality, we can focus on a principal's deviation to a profile of transfer schedules only thanks to Proposition \ref{prop:dev_to_transfer_schedules}. Upon principal $m$'s deviaton to any profile of transfer schedules, he cannot get a payoff greater than $ \underline{V}^{m}$ when the other principals responds with $\left\{ \tilde{t}^{m,k}\right\} _{k\in \mathcal{M}\backslash \{m\}}$ and the agents choose the optimal action choice for principal $m$ if there are multiple optimal action choices.

A key difference of the static version of a folk theorem is that players instantly punish a deviating principal and therefore, a folk theorem can be estalished independent of the repeated game effect. Folk theorems in static games are not new. For example, Yamashita (2010) provides a folk theorem in a static game but the set of equilibrium allocations is not well-defined because they are tied to the set of mechanisms allowed in a game. When a competing mechanism game allows very complex mechanisms, it is hard to characterize the set of equilibrium allocations in terms of model primitives. In contrast, our folk theorem characterizes the set of equilibrium allocation in terms of model primitives, regardless of which mechanisms are allowed in the game. Furthermore, the implementation of an equilibrium allocation is simple because of the tractability of DRMs.

Because a truthful $\Lambda$-equilibrium allocation requires only individual rationality for principals subject to agents' optimal action choices, the set of truthful $\Lambda $-equilibrium allocation is large. For example, principals can reach an efficient truthful $\Lambda$-equilibrium but also a truthful $\Lambda$-equilibrium where their payoffs are jointly maximized. Suppose that for joint profit maximization, sellers can tacitly collude for monopoly prices when no seller deviates. This collusion can be sustained when they use DRMs where they charge the monopoly price when agents report no deviation by a competing seller but they lower the price down to their cost when agents report a competing seller. Importantly, sellers can charge a monopoly price just by letting buyers report a deviation by a competing seller without asking them to bring hard evidence or relying on repeated game effects.

As a robustness check on our equilibrium analysis, Appendix \ref{sec:CMGPTA_Gamma} considers more complex mechanisms than DRMs such that messages available in a mechanism are arbitrarily complex and transfer schedules for agents depend not only on their messages, but also on the principal's message to himself. Theorem \ref{thm:regular} shows that the set of equilibra with more complex mechanisms is the same as the set of PIR-AM allocations. Because the set of PIR-AM allocations is the set of truthful $\Lambda$-equilibrium allocations, there is no loss of generality to consider only DRMs rather than a broader set of complex mechanisms.

\subsection{Agents with Two Actions and No Direct Preferences \label{2by2}}

Principals can support a large set of allocations in truthful $\Lambda $-equilibrium using DRMs. All $\Lambda $-equilibrium allocations are supported by agents' truthful reporting when a principal deviates. However, our static folk theorem does not predict whether agents actually truthfully report a deviation by a principal or what equilibrium allocation would emerge. To see this through lab experiments, this section construct games with 2 principals and 2 agents, where agents have only binary actions and care solely about monetary payoffs ($F_{n}(s_{n})=0\ \forall s_{n}\in S_{n}, n \in \mathcal{N}$). When there are 2 agents, the majority reporting rule that assigns the deviator-punishing transfer schedule in DRMs is equivalent to a unanimity rule. That is, the zero transfer schedule is offered whenever there is any disagreement between the agents' reports.

Distinguish principals as Principal 1 and 2 and agents as Agent A and B. Agent A's actions are Up ($U$) and Down ($D$), while B's are Left ($L$) and Right ($R$). The gross payoff matrix (before monetary transfers) for principals from the game can be written as in Figure \ref{fig:example_matrices} below.
\begin{figure}[H]
\centering
\par
\begin{subfigure}[b]{\textwidth}
    \[
    \begin{array}{c|cc}
     & \text{L} & \text{R} \\
    \hline
    \text{U} & (x_1,y_1) & (x_2,y_2) \\
    \text{D} & (x_3,y_3) & (x_4,y_4) \\
    \end{array}
    \]
  \end{subfigure}
\caption{Example 2 $\times$ 2 Game Matrix}
\label{fig:example_matrices}
\end{figure}
In the above game, $x$ payoffs represent Principal 1's gross earnings for a particular outcome and $y$ payoffs represent Principal 2's gross payoffs, respectively. Denote by $\underline{G}^{m}$ and $\overline{G}^{m}$ the minimum and maximum gross payoffs, so that $\underline{G}^{1}=\min \{x_{1},x_{2},x_{3},x_{4}\}$ and $\underline{G}^{2}=\min \{y_{1},y_{2},y_{3},y_{4}\}$, and $\overline{G}^1=\max\{x_1, x_2, x_3, x_4\}$ and $\overline{G}^2=\max\{y_1, y_2, y_3, y_4\}$. We assume the values refer to surplus so that $\underline{G}^{m}$ is non-negative.

Recall that the set of PIR-AM allocations $Z^{\ast}$ is the set of truthful $\Lambda$-equilibrium allocations and that the lower bound of each principal $m$'s equilibrium payoff is $\underline{V}^{m}$. In the games with 2 principals and 2 agents considered in this section, this lower bound becomes $\underline{G}^{m},$  the minimum gross payoff (before monetary transfers), since principals can always offer zero for all actions and guarantee $\underline{G}^m$. Therefore, any action pair can be supported as an equilibrium action pair. We redefine the condition PIR as follows: An allocation $(s,d)\in S\times  \mathbb{R}_{+}^{2\times 2}$ satisfies the condition PIR if
\begin{equation}
G^{m}(s)-\sum_{n\in \mathcal{N}}d_{n}^{m}\geq \underline{G}^{m},\forall \
m\in \mathcal{M}\text{.}  \label{new_PIR}
\end{equation}

\begin{proposition}
\label{prop:2by2}The set of PIR-AM allocations that satisfy (\ref{AM}) and (\ref{new_PIR}) is the set of truthful $\Lambda$-equilibrium allocations.
\end{proposition}

\begin{proof}
See Appendix \ref{sec:pf:prop2by2}.
\end{proof}

\section{Experimental Design\label{sec:design}}

We designed a lab experiment to empirically study our theoretical model in a simplified 2 $\times$ 2 $\times$ 2 environment from Sec \ref{2by2}. In this setting, agents care only about monetary transfers and have no intrinsic preference over outcomes. The design captures the core strategic dynamics of the theory, especially agents' ability to tacitly collude by sending false reports to principals. By observing how participants respond to DRMs in repeated interactions, the experiment allows us to examine coordination challenges, dynamic learning, and efficiency in realized outcomes.

Operationalizing our DRM in the lab allows us an opportunity to study agent reporting coordination. In particular, we examine whether agents tacitly collude on their reports when it is profitable and how well they coordinate their reports as play progresses. Our design also provides data for behavior under two scenarios for principals: one in which computerized agents always report the truth, and one in which human agents are free to send any report, including collusion on false reports. Unlike with human agents, there is no possibility of collusion with computerized agents, though principals may still attempt to collude on monopoly rents.

\subsection{Experimental Procedure}

Each experimental session involved multiple fixed groups of four participants. Two group members were randomly assigned to act as principals (Bidder 1 and Bidder 2) and the other two as agents (Row Agent and Column Agent). Participants remained in the same group throughout the session to allow learning and coordination. Several design features are borrowed from Ensthaler et al. (2020) and adapted for our study.

Each principal designed offers using a DRM, which included two transfer schedules, labeled A and B. Prior to agent reporting, principals elected to either stay with their DRM or submit a new, single-transfer schedule, labeled C. Screenshots of what principals saw at these two stages are in Figures \ref{fig:offersAB},  \ref{fig:devchoiceshot}, and \ref{fig:offersC} found in the Appendix.

Agents observe the offers made by all principals, which is either the DRM or the single transfer schedule. They then report to each principal whether the other principal deviated from their DRM. The reporting screen they saw can be found in Figure \ref{fig:reports} in the Appendix. For principals staying with the DRM, schedule A was implemented if both agents reported that the other principal did not deviate, while schedule B was implemented if both reported that the other principal did deviate. If the agents' reports disagreed, the zero transfer schedule was triggered, offering no payment to agents regardless of their actions. If principals made new offers in the
single transfer schedule, these offers are automatically triggered regardless of messages\footnote{Recall that upon deviation to a single transfer schedule, messages no longer play a role in determining action-contingent offers.}. Figures \ref{fig:submitted} and \ref{fig:action} in the Appendix shows the screen with final offers.

Sessions were run under two treatments. The first treatment uses computer agents who always reported truthfully and chose the action that maximized their payments as in Ensthaler et al. (2020). In these sessions, participants played only as principals. This treatment allows us to observe collusion by principals when agents behave exactly as in our theory and is used as a baseline for mechanism performance. The second and main treatment uses human agents, where participants acted as a principal or an agent, and agents could report strategically. In these sessions, agents were explicitly informed that their reports influenced which transfer schedule was implemented. By removing the possibility of strategic misreporting by agents in the computer agent treatment, our design allows us to cleanly attribute any observed deviations in principal offer behavior in the human agent treatment to the introduction of strategic uncertainty, coordination challenges, and the potential for collusion among agents.

Each session features two payoff matrices denoted by G1 and G2, with the game matrix changing halfway into the session. To control for order effects, some sessions started with G1 in rounds 1-8
while others started with G2. As in Section \ref{2by2}, agents could choose one of two actions: the Row Agent chose between Up (U) and Down (D), and the Column Agent between Left (L) and Right (R). These actions, combined, determined the outcome in a 2 $\times$ 2 payoff matrix shown in Figure \ref{fig:payoff_matrices}. Principals earned points from the payoff matrix of the relevant game determined by the outcome. They then transfer any payments promised to agents for those actions.

\begin{figure}[H]
\centering
\par
\begin{subfigure}[b]{0.4\textwidth}
    \[
    \begin{array}{c|cc}
     & \text{L} & \text{R} \\
    \hline
    \text{U} & (40,40) & (0,60) \\
    \text{D} & (60,0) & (10,10) \\
    \end{array}
    \]
    \caption{G1: UL Efficient}
  \end{subfigure}
\quad
\begin{subfigure}[b]{0.4\textwidth}
    \[
    \begin{array}{c|cc}
     & \text{L} & \text{R} \\
    \hline
    \text{U} & (40,40) & (0,90) \\
    \text{D} & (90,0) & (10,10) \\
    \end{array}
    \]
    \caption{G2: UR/DL Efficient}
  \end{subfigure}
\par
\vspace{\baselineskip} {\footnotesize \ \textit{Note: For each outcome, the
left value is earned by Principal (Bidder) 1 and the right by Principal
(Bidder) 2. Subjects in 2 of the 4 computer agents sessions played G1 in
rounds 1-8 and then G2 in rounds 9-16, while the other 2 sessions played G2
first. Subjects in 3 of the 7 Human Agents sessions played G1 in rounds 1-8
and then G2 in rounds 9-16, while the other 4 sessions played G2 first.} }
\caption{Experiment Payoff Matrices}
\label{fig:payoff_matrices}
\end{figure}

Importantly, human agents could benefit from coordination on false reports. For example, if both reported that a principal deviated - even if they had not - they could trigger the more generous schedule B if offers in B are higher. To the extent that these extra profits are non-negligible, agents may risk sending mixed reports and tacitly collude on a false message. Thus, the experimental setting created endogenous opportunities for tacit collusion on false reports, but also punished miscoordination via the zero transfer schedule, introducing a high-stakes coordination problem with strategic considerations aligned with our theoretical model.

The timing of the game is given by Figure \ref{fig:game_flowchart}. Bidders observe G1 or G2 and enter offer amounts for U, D, L, and R in each of two transfer schedule profiles, labeled A and B
for their DRM, for 8 total offers. In the computer agent treatment with only truthful reports, there are no mixed reports, and final transfer schedules are determined directly from both bidder's deviation choices. In the human agent sessions, reports determine offers and inconsistent reports are possible. When agents' reports are inconsistent, the DRM assigns the zero transfer schedule profile which offers zero regardless of agent's action choices. Agents then observe the final action-contingent offers and choose their actions. The pair of actions constitutes an outcome, which assigns points to principals, and transfers from principals to agents are made given the actions of each agent.

\begin{figure}[H]
\centering
\begin{tikzpicture}[
        node distance=2cm,
        every node/.style={draw, rounded corners, align=center, text width=14cm} % Adjust the text width as needed
    ]
        % Nodes
        \node (start) {Bidders observe G1 or G2 and make offers in schedules A and B for their respective DRMs};
        \node [below of=start] (player1) {Each bidder observes schedules A and B for both bidders and makes a deviation choice};
        \node [below of=player1] (player2) {Agents observe submitted offers in the DRM (schedules A and B) or single transfer schedule (C) for each bidder};
        \node [below of=player2] (outcomes) {Agents send a binary report to each bidder about the other bidder's deviation choice to determine final transfer schedule};
        \node [below of=outcomes] (end) {Agents observe offers, choose actions, and outcomes and payoffs are realized};

        % Arrows
        \draw[->] (start) -- (player1);
        \draw[->] (player1) -- (player2);
        \draw[->] (player2) -- (outcomes);
        \draw[->] (outcomes) -- (end);
    \end{tikzpicture}
\caption{Timing of the DRM Experiments}
\label{fig:game_flowchart}
\end{figure}

It is worth mentioning that in our experiment, a bidder can deviate to a single transfer schedule profile only after observing the other bidder's DRM, whereas in our theory, a bidder can choose either a DRM or a single
transfer schedule profile simultaneously. However, both games produce the same set of equilibrium allocations because, in our theory, a profile of DRMs is an equilibrium mechanism when each bidder has no incentive to
deviate to a single transfer schedule profile given the other bidder's DRM. We choose the sequential structure for bidders in order to maintain and present the common knowledge assumption on the other bidder's mechanism on the
equilibrium path in the simplest way to participants.

To limit losses that occur when an outcome costs more than it earns, bidders are endowed a budget of 100 points in each round. Agents receive this endowment as well, but do not make bids. After observing the other bidder's
tentative DRM as well as their own, bidders decide to stay with the DRM or offer a new, single transfer schedule profile C, and enter new offer amounts for each of the four actions if they choose to deviate. Agents then observe
final offers and whether a bidder deviates from their DRM and sends a binary report to the other bidder about whether that bidder deviated. Crucially, bidders never observe the final offers made to agents by other bidders and receive only the reports from agents. After reporting, final offers are made, agents choose their action, and transfers and payoffs are realized accordingly. In the computer agent treatment, in the event of equal offers for
each action, the action is chosen randomly with equal probability. Participants are told that points earned by computer agents are not paid out.

Given the complexity of the experimental environment, 3 practice rounds were
played before groups were reshuffled and fixed for 16 payment rounds. Roles
were maintained between the practice and payment rounds. A brief
comprehension quiz was also included prior to the practice rounds and paid
subjects 8 experimental points (\$0.56 CAD) for correct responses to improve
their understanding and attention (Freeman et al., 2018). Following quiz responses,
descriptions for arriving at the correct answer for each comprehension
question was provided, and any incorrect response had to be corrected before
proceeding to ensure the descriptions are read. Subjects begin the payment
rounds with payoff matrix G1 or G2 for 8 rounds each, changing at round 9.

We conducted 4 sessions of the Computer agents treatment with 74 subjects
and 7 sessions of the Human agents treatment with 140 subjects, for 214
total subjects and 37 and 35\footnote{One group's messages were not recorded, leaving 34 groups available for
study.} groups for the computer agent and human agent treatments,
respectively. Demographic information for our sample can be found in the
Appendix. Sessions with Computer agents lasted approximately 90 minutes, while
sessions with human agents \footnote{The first 4 sessions of the human agent treatment featured a display error
where Bidder 2's payoffs were displayed incorrectly. An announcement was
made during the sessions regarding the display and that points were
calculated based on earnings from the game matrix minus transfer amounts,
despite the displayed payoff not adding this correctly. An indicator
variable equal to 1 for sessions 1-4 in the human agent treatment and 0 for
sessions 5-7 is included in the regression analysis and is always
insignificant with large p-values.} lasted approximately 2 hours. All sessions were
held at the McMaster Decision Sciences Laboratory at McMaster University.
All experiments were programmed with o-Tree (Chen et al., 2016). Payoffs
were earned as points in the experiment and converted to CAD at a rate of 7
cents CAD per point. Three rounds from the 16 paid rounds were randomly
chosen with equal probability for payment. Subjects earned on average
\$30.51 CAD including a \$5 CAD show-up payment.

\section{Results\label{sec:results}}

The results from our experiment are organized by agent and principal behavior. Our main experimental results concern agents' reporting behavior in the human agents sessions and are presented first. We report the frequency of successful message coordination across rounds. We then disaggregated message pairs into three types - double true reports, double false reports, and mixed reports - to see whether true reports or false reports drive the changes across time. To understand the factors involved in successful tacit collusion, we estimate a random-effects model controlling for group heterogeneity. In addition, we report our findings on outcomes, offers, and deviations and compare them with the computer agent sessions. Since agents always chose the action that earned them the highest points, satisfying AM in both types of sessions, we refer to these results as principal results. Our results for principals provide evidence on behavior under two settings: one in which agents have no possibility of collusion (computers), and one in which agents may report strategically (humans).

\subsection{Agents}

Our experimental data contain 1,088 message pairs (16 rounds $\times$ 2 message
pairs $\times$ 34 groups). From these message pairs, 724 are sent to non-deviating
principals, leaving us with 724 meaningful message pairs for analysis\footnote{Note that while reports to a deviating principal may be false, they do not determine offers and therefore false reports sent a principal who deviated are not meaningfully tied to incentives.}.

Analysis of the message pairs shows two sides of the story. A majority of
the time, agents do truthfully report. Of all message pairs, approximately
67.3\% are pairs of truthful reports. This seems to support the validity of
equilibrium characterization of static games based on truthful reporting in
competing mechanism games (or partial implementation with a single
principal's mechanism design). Looking at individuals more closely reveals
that 29.41\% (20/68 agents) always send truthful reports, and another 14.7\%
(10/68) falsely reported less than 10\% of the time.

\begin{table}[]
\centering
\resizebox{\textwidth}{!}{
\begin{tabular}{ccccc}
\hline
 \textbf{Groups} & \textbf{Both True} & \textbf{Both Lie} & \textbf{Mixed} \\ \hline
Percent of Aggregate & 67.27 & 6.63 &  26.10 \\ \hline
 & & & & \\ \hline
 \textbf{Individuals} & \textbf{Never Lie} & \textbf{Rarely Lie (under 10\%)} & \textbf{Some Lies (under 25\%)} & \textbf{Common Lies (25\%+)} \\ \hline
Percent of Aggregate & 29.41 & 14.71 & 22.06 & 33.82 \\ \hline
\end{tabular}
}
\caption{Agent's Reporting}
\label{tab:Agents_reporting}
\end{table}

\begin{figure}[H]
\centering
\includegraphics[width=.5\textwidth]{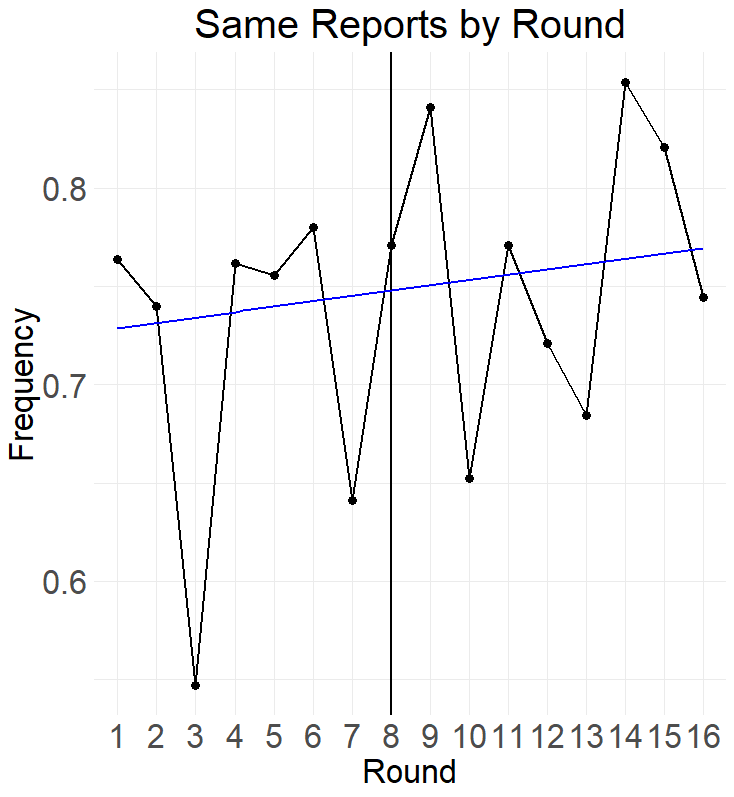}
\caption{Numbers of Consistent Message Pairs by Round}
\label{fig:same_reports}
\end{figure}

As rounds progress, agents learn to play with their counterparty agent and that mixed reports result in low payoffs. Coordination on reports improved as time passed, with the number of mixed report pairs decreasing with rounds. Figure \ref{fig:same_reports} plots the proportion of message pairs that are the same, either both true or both false.

\begin{figure}[H]
\centering
\includegraphics[width=1\textwidth]{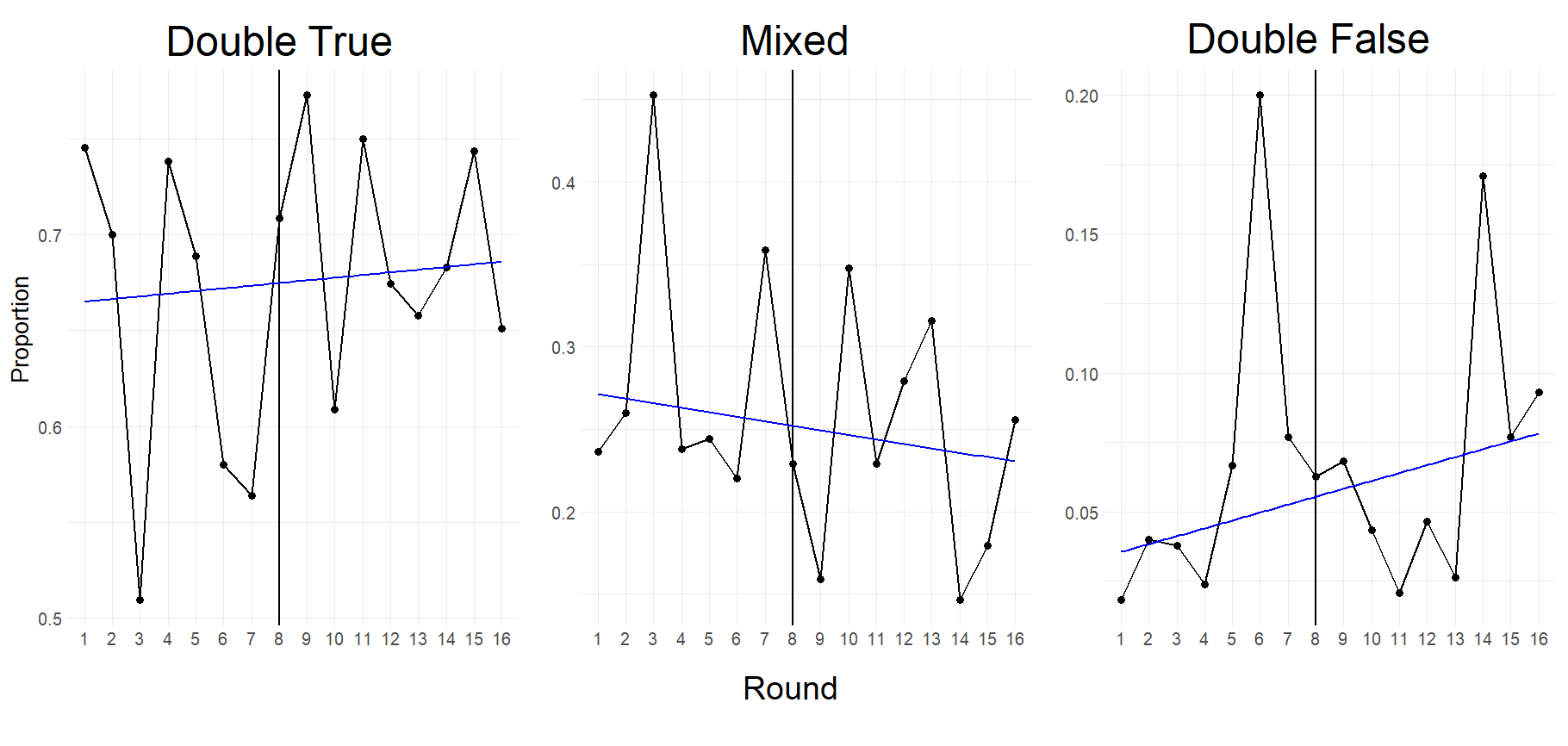}
\caption{Numbers of Message Pairs by Type across Rounds}
\label{fig: reports_by_type}
\end{figure}

Figure \ref{fig: reports_by_type} separately plots the number of message pairs by type, either both true, both false, or mixed, across rounds. The frequency of mixed reports decreases over time, driven by slight increases in truthful reports and moderate increases in false reports.

To understand the relationship between groups' false reporting and incentives, we estimate a random-effects Logit model. Random effects are at the group level. The specification for our regression model is as follows:

\begin{eqnarray}
\text{Y}_{it} &=&\beta _{0}+\beta _{1}\times \text{Both Incentive to Lie}%
_{it}+\beta _{2}\times \text{ Round}_{t}  \label{Logit} \\
&+&\beta _{3}\times \text{Other Bidder Deviated}_{it}+\beta _{4}\times \text{G1
First}_{it}+\epsilon _{it},  \notag \\
\epsilon _{it} &=&\alpha _{i}+u_{it}  \notag
\end{eqnarray}

where \textit{Y} is an indicator for double false reports (Both Lie) or
double true reports (Both True), \textit{i} indexes groups, and \textit{t}
indexes rounds. The variable

The variable \textit{Both Incentive to Lie} is a binary variable equal to 1 when both agents have an incentive to send a false report. This is calculated by first finding the maximum offer between both transfer schedules of a non-deviating bidder relevant to the particular agent. If the maximum offer relevant to an agent is in schedule profile B from some non-deviating bidder, then the agent has an incentive to report to the bidder that the other bidder deviated, whether they have or not, to implement B. Thus, in the case where an agent reports to a non-deviating bidder with a maximum offer in Schedule profile B, their incentive to lie indicator variable is equal to 1 at the individual level when the other bidder did not deviate. Both agents can falsely report a deviation and implement B. Similarly, in the case where an agent reports to a non-deviating bidder with a maximum offer in schedule profile A, their incentive to lie indicator variable is equal to 1 at the individual level when the other bidder deviates. In this case, both agents can falsely report that the deviating bidder stayed with the original DRM, implementing A. In other cases, such as where the maximum offer is in schedule profile B but the other bidder did indeed deviate, the individual incentive to lie is equal to 0, since false reports implement B but the highest offer is in A. Given that both agents must send the same report to receive non-zero offers, we combine the individual incentives and set \textit{Both Incentive to Lie} equal to 1 when both individual incentives are equal to 1.

The \textit{Other Bidder Deviated} variable captures the effect of falsely reporting a deviating or non-deviating bidder. It is equal to 1 when the bidder whom the message is about deviates, and 0 when they do not deviate. A positive coefficient suggests that groups are more likely to coordinate on a false report when reporting about a deviating bidder than falsely report about a non-deviating bidder. \textit{G1 First} captures the effect of playing G1 in the first 8 rounds on the propensity for both agents to falsely report compared to those playing G2 in the first 8 rounds. The variable \textit{Both agents Female} is an indicator for when both the Row Agent and Column Agent of a particular group are female. In addition, we run a version with independent variable \textit{Both Incentive Size} instead of the incentive indicator, which is the difference in the maximum amount of points an individual agent can earn from reporting double false reports and the maximum they can earn from reporting truthfully. This variable takes a positive value when the combined incentive size for both agents is positive. A positive value indicates a joint incentive to lie, with the magnitude of the amount representing the amount of the incentive. Table \ref{tab:logit_estimation} shows the regression results.

\begin{table}[H]
\caption{{}Random-Effects Logit }
\label{tab:logit_estimation}\centering
\centering
\begin{tabular}{@{\extracolsep{5pt}}lcccc}
&  &  &  &  \\[-1.8ex] \hline\hline
&  &  &  &  \\[-1.8ex]
& \multicolumn{4}{c}{\textit{Dependent variable:}} \\ \cline{2-5}
&  &  &  &  \\[-1.8ex]
& \multicolumn{2}{c}{Both Lie} & \multicolumn{2}{c}{Both True} \\
&  &  &  &  \\[-1.8ex]
& (1) & (2) & (3) & (4) \\ \hline
&  &  &  &  \\[-1.8ex]
Both Incentive to Lie & 1.485$^{***}$ &  & $-$0.716$^{***}$ &  \\
& (0.398) &  & (0.260) &  \\
&  &  &  &  \\
Both Incentive Size &  & 0.050$^{***}$ &  & $-$0.037$^{***}$ \\
&  & (0.013) &  & (0.012) \\
&  &  &  &  \\
Round & 0.067$^{*}$ & 0.075$^{**}$ & 0.010 & 0.006 \\
& (0.037) & (0.037) & (0.020) & (0.020) \\
&  &  &  &  \\
Both Agents Female & $-$1.094 & $-$1.069 & 0.669 & 0.646 \\
& (0.849) & (0.871) & (0.566) & (0.562) \\
&  &  &  &  \\
Other Dev & $-$0.075 & 0.064 & 0.303 & 0.263 \\
& (0.394) & (0.401) & (0.221) & (0.222) \\
&  &  &  &  \\
Soc Eff First & $-$0.407 & $-$0.322 & 0.174 & 0.141 \\
& (1.043) & (1.065) & (0.757) & (0.751) \\
&  &  &  &  \\
First 4 & $-$0.147 & $-$0.092 & 0.013 & 0.023 \\
& (1.097) & (1.121) & (0.801) & (0.796) \\
&  &  &  &  \\
Constant & $-$3.952$^{***}$ & $-$4.108$^{***}$ & 0.565 & 0.625 \\
& (0.945) & (0.984) & (0.582) & (0.581) \\
&  &  &  & \textit{Note:} \\ \hline
&  &  &  &  \\[-1.8ex]
\textit{Note:} & \multicolumn{4}{r}{$^{*}$p$<$0.1; $^{**}$p$<$0.05; $^{***}$p%
$<$0.01} \\
&  &  &  &
\end{tabular}%
\end{table}

Incentives have a statistically significant effect on the probability of double false reports. The presence of an incentive for both agents to falsely report, captured by an indicator variable in specifications (1) and (3), increases the probability of double false reports by 7.4\% on average. Considering the size of the incentive instead in specifications (2) and (4), a 1 point incentive size for the group, equivalent to 8 cents CAD, increases the probability of double false reports by 0.245\%. This sensitivity to the size of the incentive is suggestive that individuals respond to the magnitude of the incentive rather than simply whether they have an incentive or not. Intuitively, since false reports are difficult to anticipate and risk coordination failure, the size of the incentive matters in taking this risk.

Each passing round increases the probability of double false reports, holding other factors fixed, with an average marginal effect of 0.34\%. This contrasts with Abeler et al. (2019) where observed reporting behaviour does not change with repeated reports, but is consistent with Kocher et al. (2018), who find that groups exhibit
increasing dishonesty over time, and with Kroher and Wolbring (2015), who emphasize that dishonesty is shaped through social learning in repeated interactions.

Individual group fixed effects have strong significance on coordinated
reports and much of the variation is across groups. We do not observe any
statistically significant differences attributable to groups composed of
only female agents, in contrast with previous literature showing that pairs
of females lie less than pairs of males or mixed pairs (Muehlheusser et al.,
2015). Our setting never refers explicitly to gender, and the identity of agents is not revealed, suggesting these effects arise only when gender is made salient.

\begin{figure}[]
\centering
\includegraphics[width=.5%
\textwidth]{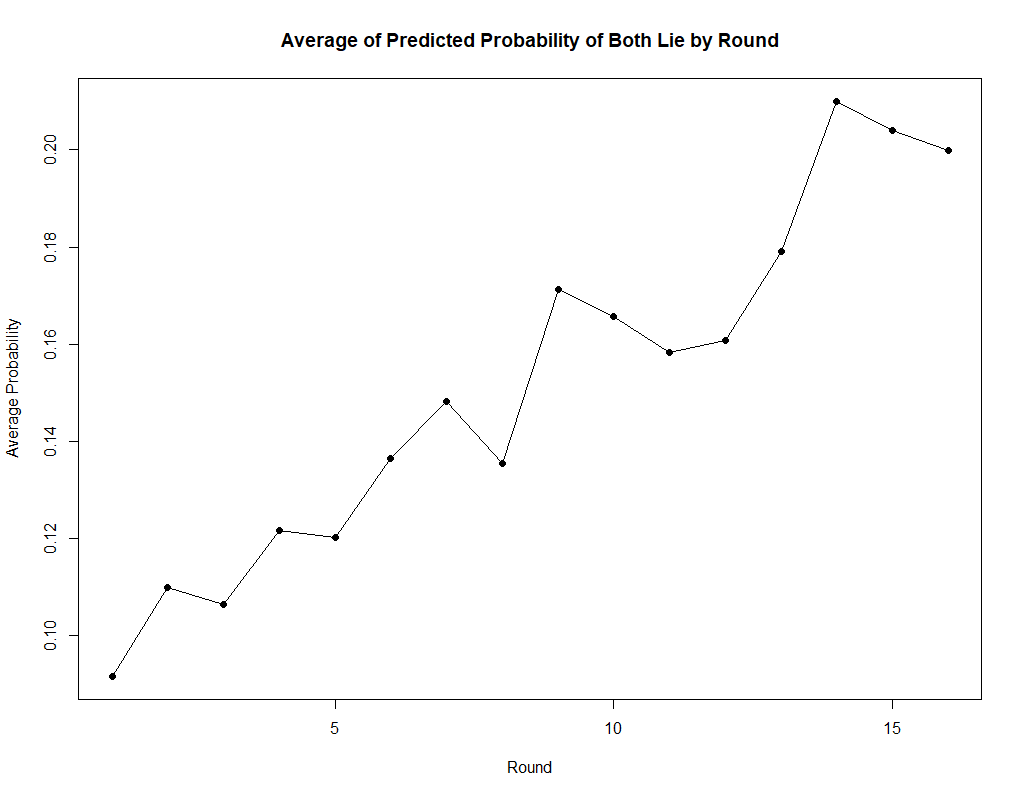}
\caption{Predicted Probability of Both Lie by Round}
\label{fig:bothlie_round}
\end{figure}

Using our model estimates, we then create a counterfactual dataset where the
incentive to lie indicator is set to 1 and plot the average predicted
probability of double false reports of all groups by round, found in
Figure \ref{fig:bothlie_round}. For these predicted amounts, we run a
reduced model of the Both Lie indicator on Both Incentive to Lie and Round,
with the statistically insignificant variables from our random-effects model omitted. Average predicted
probability of double false reports increases from under 9\% at the
beginning of the experiment to about 20\% by the end when there is a consistent
incentive to lie\footnote{Note that the non-monotonicity of the plot is due to the differential
presence of groups across rounds, since message pairs to deviating bidders
are irrelevant and not included.}. For many groups, the probability rises to
nearly 40\% by the end, while for the mostly truthful groups the probability
of double false reports remains below 5\% even at the last round. This
increase suggests some underlying learning about whether partner agents are
willing to falsely report and provides evidence on dynamic lying behavior in
group settings, of which there is relatively little.

\begin{figure}[H]
\centering
\includegraphics[width=.8\textwidth]{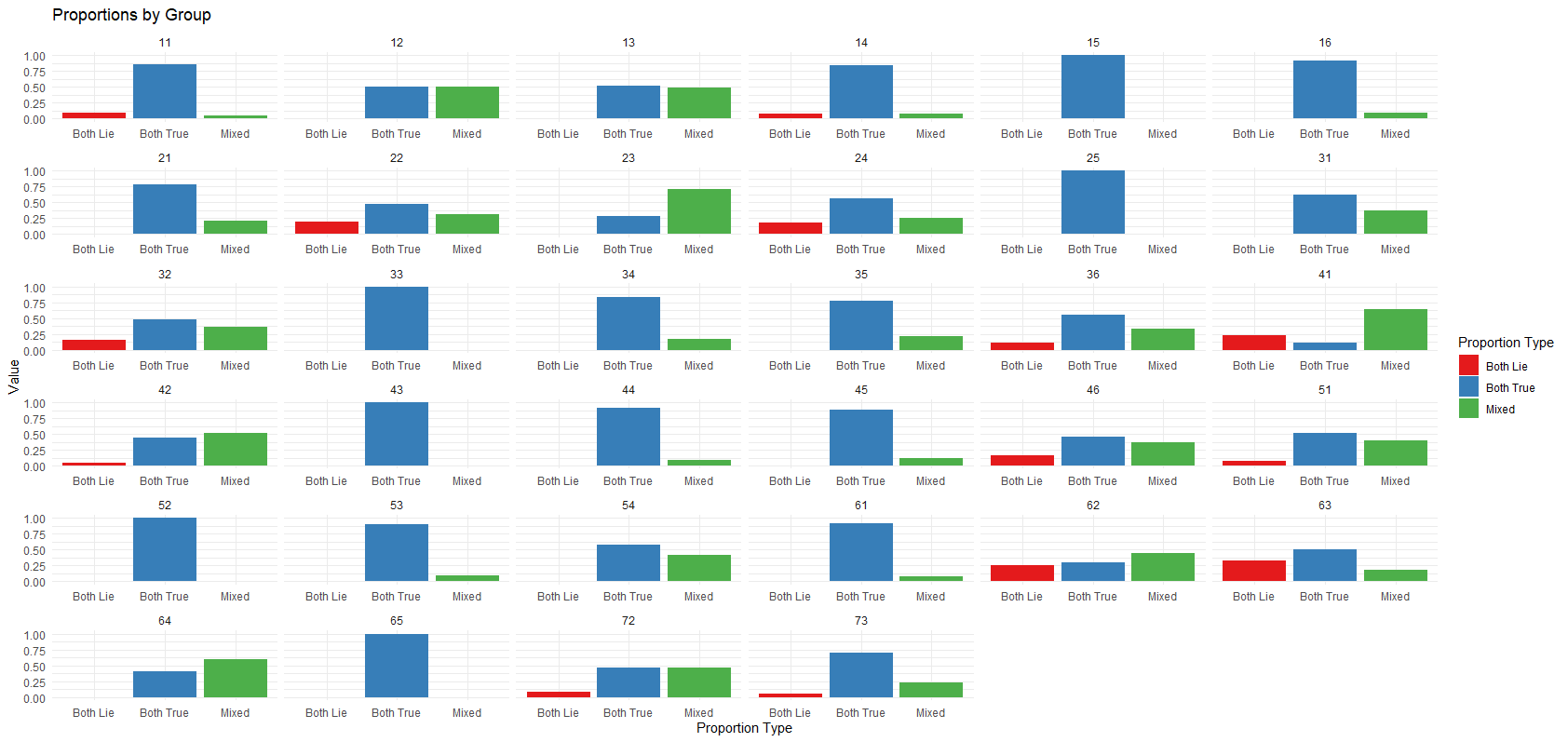}
\caption{Message Pair Types by Group across Rounds}
\label{fig:split_groups}
\end{figure}

Groups are quite heterogeneous in their reporting behavior, with groups' success at coordinating reports differing substantially. Figure \ref{fig:split_groups} plots the proportion of double true reports, double false reports, and mixed reports by group. Double true reports is the most frequent reporting outcome for most groups. As well, most groups never coordinate on double false reports. For 6 of the 34 groups, only true reports were sent. In several others, only a few false reports were sent. Groups that were able to coordinate on double false reports had substantial mixed report outcomes as well, highlighting the difficulty of this type of coordination. These mixed reports proved costly; Agents in groups who had conflicting reports less than 25\% of the time earned about 9.6\% more than agents from groups with conflicting reports 25\% or more of the time. Agents in groups with no double false reports earned about 8\% more than agents in groups with at least one double false report, highlighting that collusion does not always pay.

\subsection{Principals}

\subsubsection{Offers and Deviations}

\begin{figure}[H]
\centering
\includegraphics[width=1\textwidth]{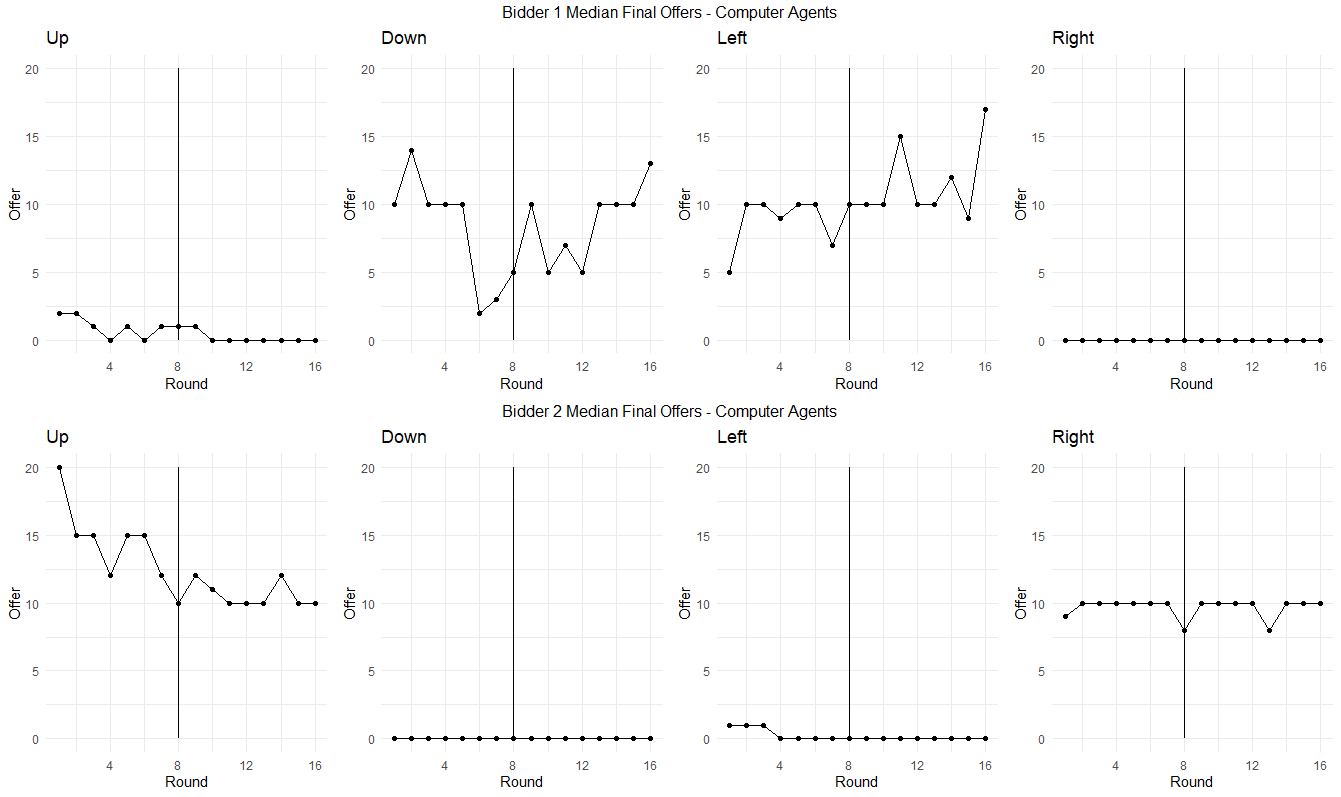}
\caption{Median Offers by Round with Computer Agents}
\label{fig:T1_med_offers}
\end{figure}

\begin{figure}[H]
\centering
\includegraphics[width=1\textwidth]{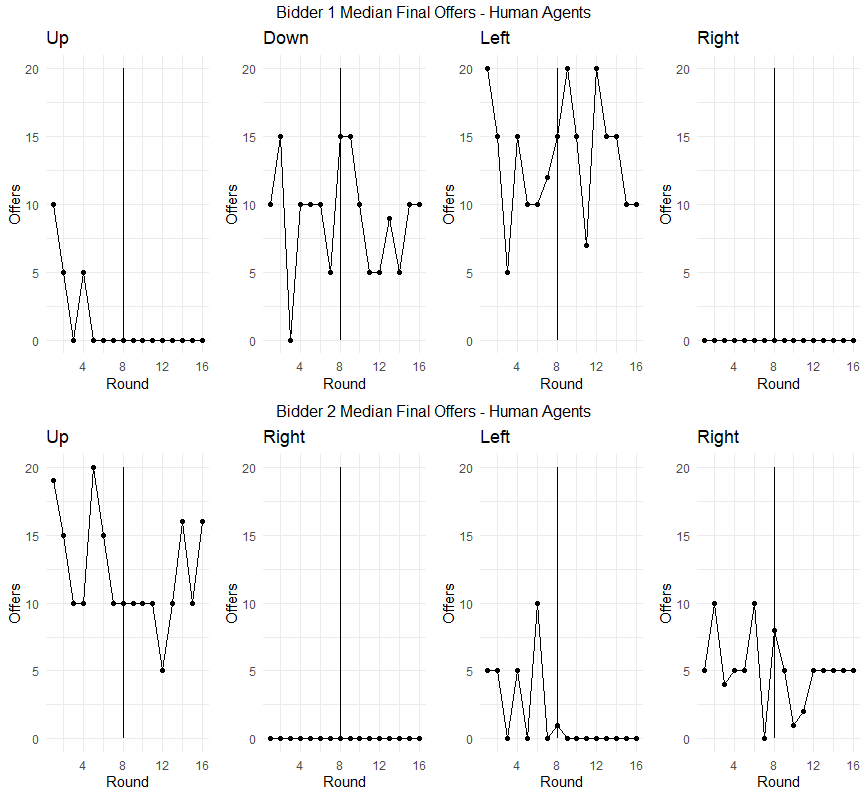}
\caption{Median Offers by Round with Human Agents}
\label{fig:T2_med_offers}
\end{figure}

The aggregate distribution of bidders' offers shows a very high frequency of 0, especially for actions D and R, with the frequency of offers mostly decreasing in offer size. Incidentally, outcome DR is the lowest paying outcome in both G1 and G2. Figures \ref{fig:T1_med_offers} and \ref{fig:T2_med_offers} show that offers tend to become competitive as play progresses. At the median, final offers for U and R go to zero for Bidder 1 while bids for D and L remain positive. The opposite is true for Bidder 2, who prefers outcome UR. In the 2 $\times$ 2 $\times$ 2 setting also considered in GPTA (Prat and Rustichini, 2003), an equilibrium condition is that principals offer only positive amounts to each Agent for their most preferred action of the two, and offer zero to each Agent for the other action. This condition is imposed by Ensthaler et al. (2020) and, while not a necessary equilibrium condition in our model, emerges organically nonetheless. This pattern held regardless of the type of agent that principals faced and for practically all groups, with offer amounts converging to about the same level by the end of the game. The somewhat high offers for L by Bidder 1 and for U by Bidder 2 is suggestive that principals were not attempting to collude on monopoly rents.

\begin{figure}[]
\centering
\includegraphics[width=1\textwidth]{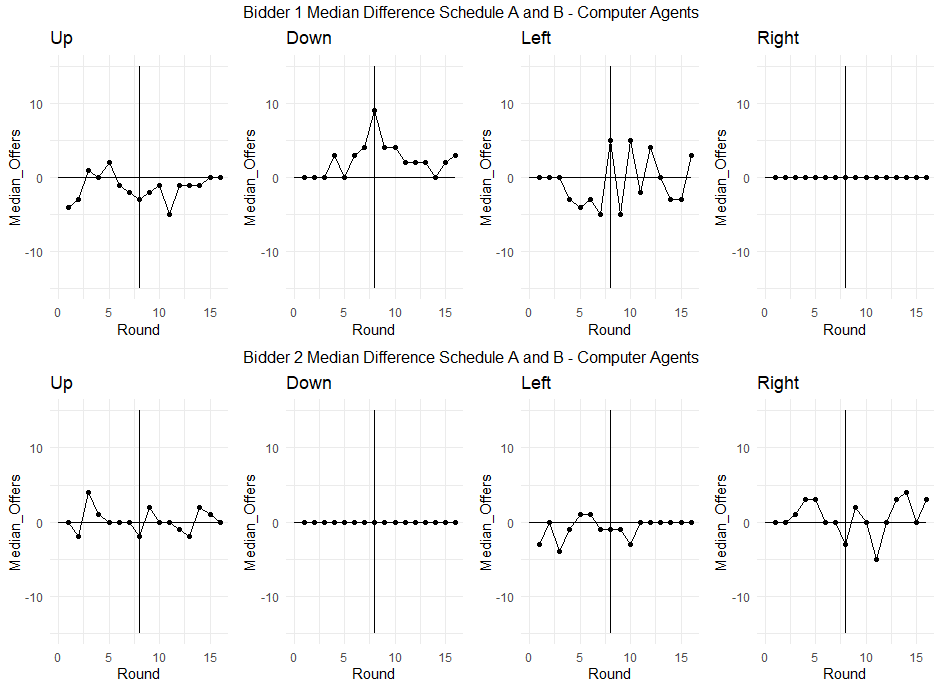}
\caption{Median Differences in offers in Schedule B minus Schedule A }
\label{fig:T1_med_diffs}
\end{figure}

Offers in general are higher for U and L than D and R\footnote{Wilcoxon Signed Rank test p $< 0.01$ comparing offers for U to D and R, and offers for L to D and R.}. In both the human and computer agent sessions, the differences between offers in schedule profiles B and A at the median are small (Appendix Figures \ref{fig:T1_med_diffs} and \ref{fig:T2_med_diffs}). These small differences can be interpreted as a defense against collusion driven by false reports, mitigating the incentive to lie. Interestingly, the game change at round 8 generally does not have a strong effect on offers, which is mostly driven by time.

We observe more deviating principals in the computer agents treatment than in the human agents treatment. About 33.8\% deviate in the human agents treatment compared to 46.3\% with computer agents. One possible explanation is the potential of getting outcomes for free in the human agents treatment when a bidder does not deviate, since human agents can send conflicting reports while computer agents never do, though we do not test this directly.

\begin{figure}[H]
\centering
\includegraphics[width=.6\textwidth]{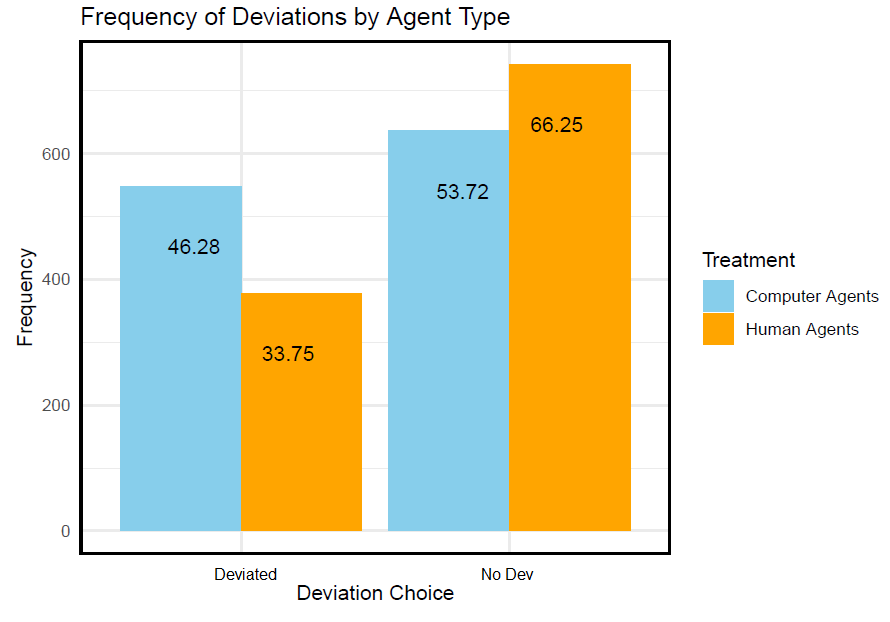}
\caption{Deviation Frequencies by Agent Type}
\label{fig:deviations}
\end{figure}

\subsubsection{Outcomes}

Although actions are ultimately selected by agents, all agents simply chose the action that offered them the highest payment, behaving according to AM. We thus report the outcomes here, following results on offers. The two games played in our experiment have different efficient outcomes. In G1, the unique efficient outcome is UL, while in G2 efficient outcomes are UR and DL. In general, implemented outcomes were efficient more often than random. In aggregate, the proportion of efficient outcomes in the Computer agents treatment is 33.1\% (p $<0.01$ compared to 25\%) for G1 and 68.2\% (p $<0.01$ compared to 50\%) in G2. We also count the number of groups where UL is the most
frequent outcome in G1 and UR or DL is the most frequent outcome in G2.\footnote{When there is a tie for most frequent outcome, each outcome is included.} From the 37 groups, UL was the most frequent outcome in G1 for 33.3\% of groups, and either UR or DL was the most frequent outcome for 71.4\% of groups. As in Ensthaler et al. (2020), we find very few instances of DR ($<8\%$ in both games), the least efficient outcome in both games.

Table \ref{tab:T1_Outcomes} provides a summary of the outcomes realized in the computer agents treatment. The distributions of outcomes are given for G1 and G2 separately. Outcomes UL, UR, and DL are implemented with similar frequencies in either game, with the relative frequency of the off-diagonal outcomes increasing and UL decreasing in G2. Individually, U or L are implemented more than 60\% of the time in G1 and approximately 60\% in G2.

\begin{table}[H]
\centering
\begin{tabular}{llll}
G1 & \multicolumn{1}{c}{\textbf{L}} & \multicolumn{1}{c}{\textbf{R}} &  \\
\multicolumn{1}{c}{\textbf{U}} & 0.33 & 0.33 & 0.66 \\
\multicolumn{1}{c}{\textbf{D}} & 0.29 & 0.05 & 0.34 \\
& 0.62 & 0.38 & 1 \\
&  &  &  \\
G2 & \multicolumn{1}{c}{\textbf{L}} & \multicolumn{1}{c}{\textbf{R}} &  \\
\multicolumn{1}{c}{\textbf{U}} & 0.25 & 0.35 & 0.60 \\
\multicolumn{1}{c}{\textbf{D}} & 0.33 & 0.07 & 0.40 \\
& 0.58 & 0.42 & 1%
\end{tabular}%
\par
\caption{Outcomes with Computer Agents}
\label{tab:T1_Outcomes}
\end{table}

Table \ref{tab:T2_Outcomes} summarizes outcomes implemented when agents are other participants. Compared to the computer agents treatment, efficiency is higher in G1 and lower in G2. With human agents, the observed proportion of efficient outcomes is 37.5\% in G1 and 61.1\% in G2. Each of these is statistically significantly different from randomness (25\% and 50\% for G1 and G2, respectively, p $<0.01$). The proportion of UL is higher in G1 than
in G2 (38\% compared to 31\%, p = 0.0166), while the proportion of UR and DL are higher in G2 than in G1 (24\% compared to 28\%, p = 0.097 and 28\% compared to 33\%, p = 0.0859, respectively). UL was the most frequent outcome for 40\% of groups in G1, and either UR or DL was the most frequent outcome for 68.9\% of groups.

\begin{table}[H]
\centering
\begin{tabular}{llll}
G1 & \multicolumn{1}{c}{\textbf{L}} & \multicolumn{1}{c}{\textbf{R}} &  \\
\multicolumn{1}{c}{\textbf{U}} & 0.38 & 0.24 & 0.62 \\
\multicolumn{1}{c}{\textbf{D}} & 0.28 & 0.11 & 0.39 \\
& 0.66 & 0.35 & 1 \\
G2 & \multicolumn{1}{c}{\textbf{L}} & \multicolumn{1}{c}{\textbf{R}} &  \\
\multicolumn{1}{c}{\textbf{U}} & 0.31 & 0.28 & 0.59 \\
\multicolumn{1}{c}{\textbf{D}} & 0.33 & 0.08 & 0.41 \\
& 0.64 & 0.36 & 1%
\end{tabular}%
\par
\caption{Outcomes with Human Agents}
\label{tab:T2_Outcomes}
\end{table}

Interestingly, we do find order effects on efficiency. When G1 is played first, efficiency is high in both games with either agent type. When G2 is played first, patterns of outcomes appear to persist when moving to rounds with G1.

\begin{figure}[H]
\centering
\includegraphics[width=%
\textwidth]{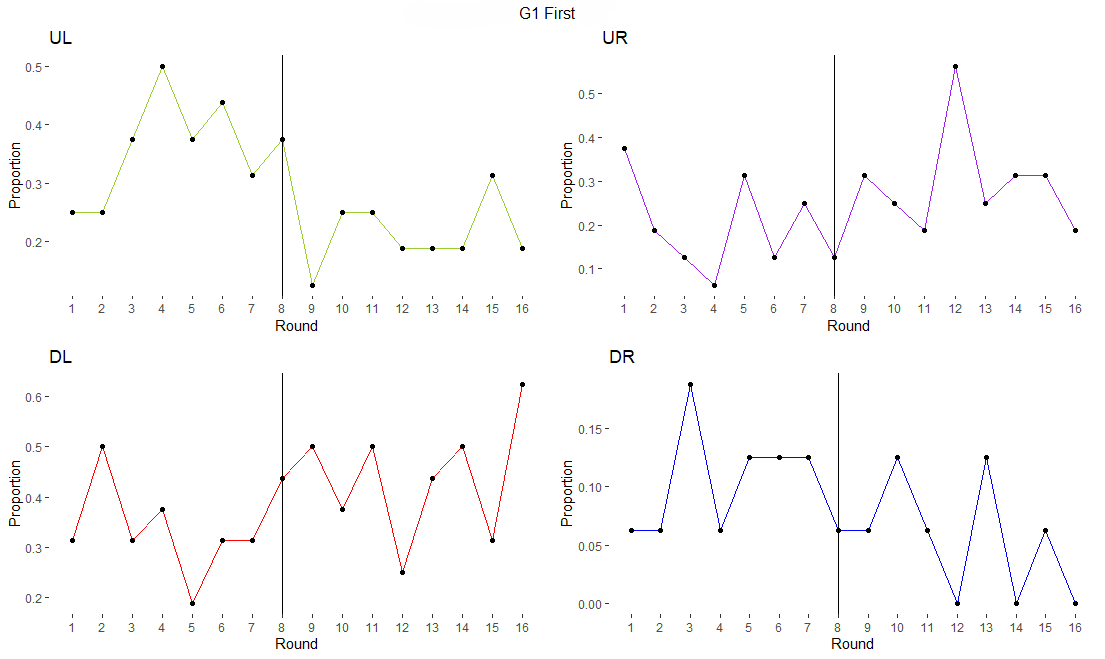} %
\includegraphics[width=%
\textwidth]{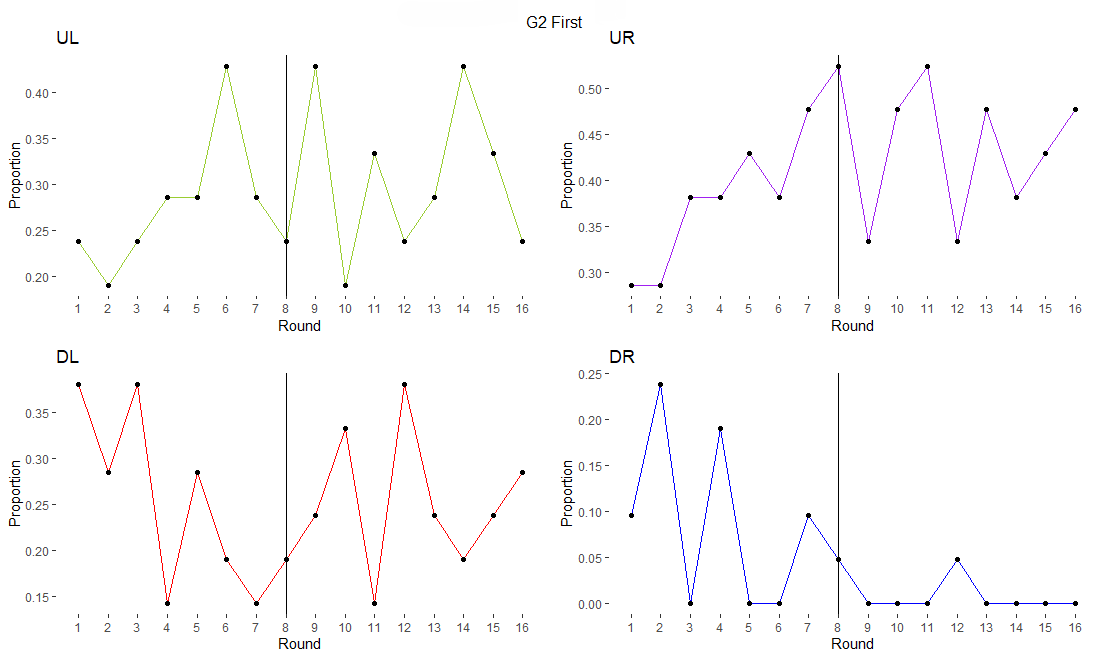}
\caption{Outcome Proportions by Round with Computer Agents}
\label{fig:T1outcomes_Rounds}
\end{figure}

Figure \ref{fig:T1outcomes_Rounds} shows the proportion of each outcome by round in separate plots for computer agents. The frequency of UL in the first 8 rounds compared to the latter 8 rounds is not statistically significantly different when G2 is played first (27.4\% compared to 31\%, p = 0.5484), but is when G1 is played first (35.9\% compared to 21.1\%, p = 0.0127). Similarly, the total proportion of UR or DL outcomes in the first 8 rounds compared to the latter 8 rounds is not statistically different when G2 is played first (64.3\% compared to 68.5\%, p = 0.4884), but is different when G1 is played first (53.9\% compared to 73.4\%, p $<0.01$). When G1 is played first, the proportion of UL is over 35\% for most rounds before dropping to under 25\% or less in
all rounds but one. Six rounds in the first half of the session had a higher frequency of UL than the highest observed frequency in the latter half. Frequencies of UR and DL increased moderately after the game switch. When G2 is played first, frequencies for UL, UR, and DL are similar after moving to G1 in round 9, though highly variable. While groups seem able to move from frequently implementing UL in G1 to UR or DL in G2, behavior is similar across games when G2 is played first. In both cases, the frequency of the least efficient outcome, DR, tends to zero.

\begin{figure}[H]
\centering
\includegraphics[width=%
\textwidth]{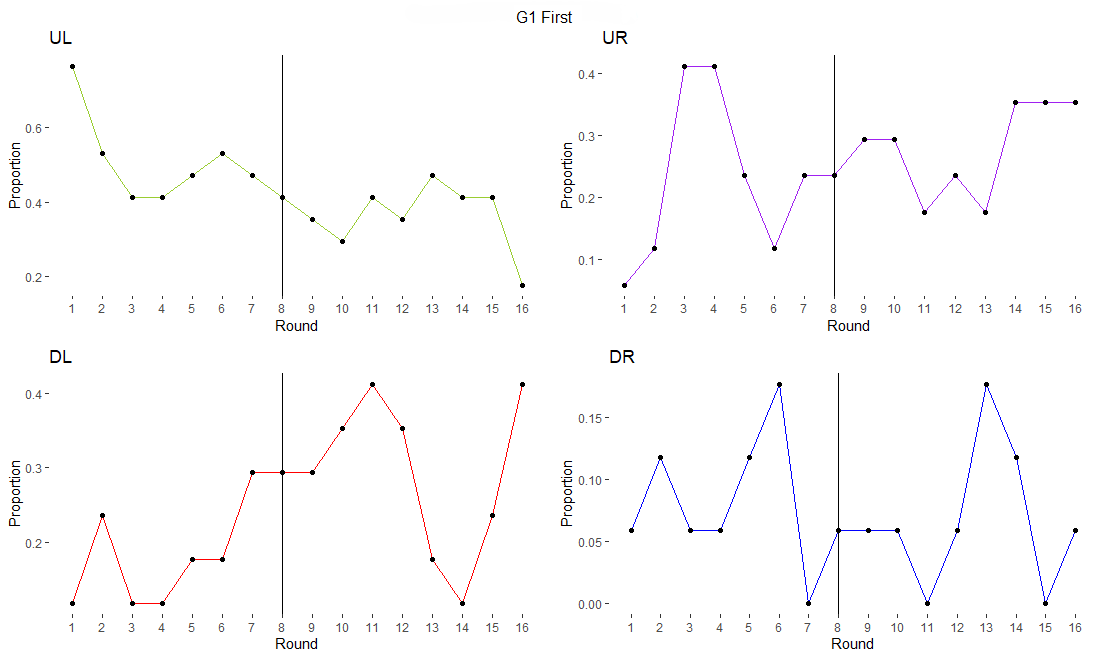} %
\includegraphics[width=%
\textwidth]{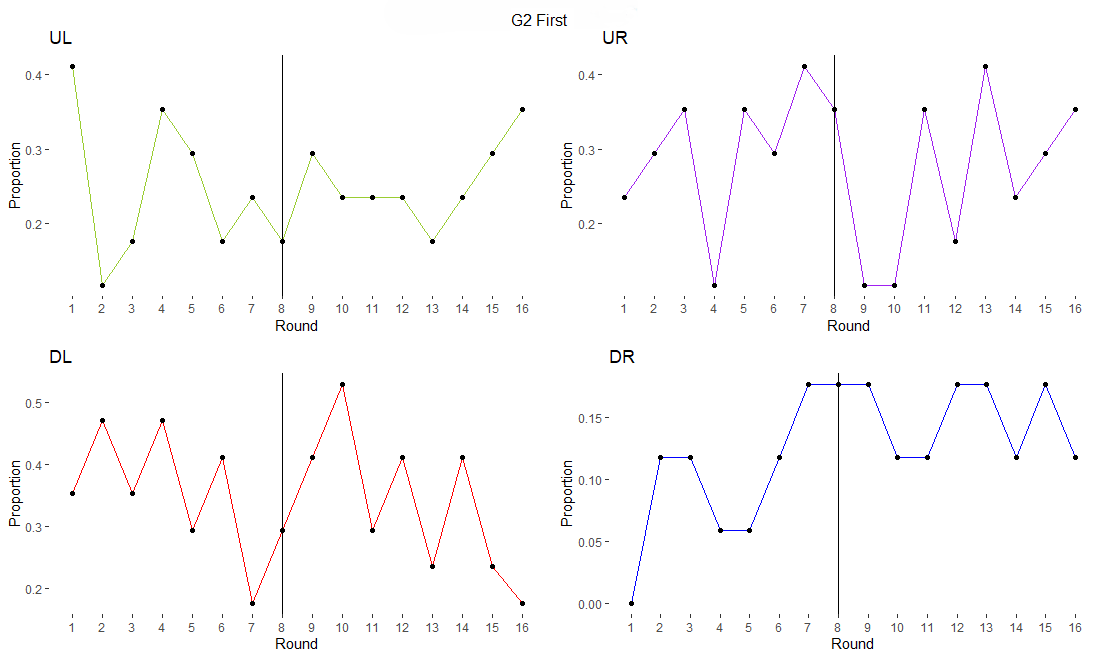}
\caption{Outcome Proportions by Round with Human Agents}
\label{fig:T2_byround}
\end{figure}

With human agents, patterns are similar to the computer agents treatment, displayed in Figure \ref{fig:T2_byround}. When G1 is played first, UL decreases in frequency from the first half of the session into the latter half (50\% compared to 36\%, p = 0.0275) with an
increase in realizing UR or DR over that time (41.9\% compared to 57.4\%, p = 0.0153). When G2 is played first, the proportion of UL outcomes remains flat (sample proportions were the same, p = 1), and there are no significant differences in the total proportion of UR and DL between the first half and the latter half (64.6\% compared to 60.4\%, p = 0.5428). Since principals' offers tended to become competitive, this behavior offers a plausible explanation for the observed order effects: when G1 is played after G2, the convergence to competitive offers has generally occurred, and implementing UL is less common than if G1 were played first, where Bidder 1 may still bid positive amounts for U and Bidder 2 may still bid positive amounts for L.

Overall, our results reveal the difficulty of coordination in the DRM game. Learning whether a partner agent is willing to send false reports coupled with the incentives to do so creates a strategically difficult coordination problem. Agents not only need to evaluate their incentives as well as their counterparty agent's incentives, and anticipate whether the counterparty agent will respond to those incentives or not. In general, truth-telling seems to be a default strategy, and therefore easier to anticipate. Some groups maintain the status-quo of truth-telling, while others frequently attempt false reports, with varied degrees of successful coordination.

Our results suggest that, in aggregate, the assumption of truthful agent reporting applies a majority of the time. Dynamically, a learning process emerged alongside the reporting coordination challenge. Most groups attempt false reporting at least once, with several of these groups reverting back to truthful reporting after coordination failure from mixed reports. If an agent learns other agents are unwilling to falsely report regardless of incentives, they themselves have no incentive to falsely report, since mixed reports always earn transfers of zero. Hence, for groups with a majority of truth-tellers, there is no incentive for any strategic type to falsely report under DRMs. Although these truth-tellers do not respond to incentives by falsely reporting, others do. Still, even with agents who are both willing to false report, it may be better for them to tell the truth, as anticipating when the other will report truthfully or not is conceptually difficult and, given the structure of DRMs, costly when reports are inconsistent across agents.

Principals' offers converge to zero for any action other than for their most preferred action. Implicitly, prinicipals then eventually never simultaneously send positive amounts for UL, the zero-inequality outcome. To
do so would risk the other principal getting their preferred outcome at a cheaper cost. In fact, in the two games we consider, principals' offers are positive only for actions that lead to the greatest \textit{inequality} - DL
for Bidder 1 and UR for Bidder 2. For agents, predicting a counterparty agent's report,
which depends on their propensity to falsely report, the received offers
from principals in schedules A and B, and their beliefs about each other, is
a difficult task. Reporting coordination improved over time. Our conjecture
is that the most important element of this improvement is learning about
one's counterparty agent, aligned with previous experimental literature, though we did not reshuffle groups throughout the experiment and therefore cannot disentangle this learning from learning about the DRM environment itself\footnote{Furthermore, a significant portion of the experiment was spent with instructions, quizzes and practice to ensure the DRM was understood before learning anything about a counterparty agent's behavior.}. Although coordination is difficult in this environment, outcomes are tend to be efficient as principals can generally avoid the worst outcome, DR. A large majority of report pairs were double
true reports, highlighting that it is this coordination difficulty that
allows the DRM to elicit the truth most of the time.

\section{Conclusion\label{sec:conclusion}}

We propose CMGPTAs, an extension of the GPTA (Prat and Rustichini
(2003)), where a principal can offer any arbitrary mechanism that specifies
a transfer schedule for each agent conditional on all agents' messages. This allows a principal to make their terms of trade responsive to the unobservable terms of trade offered by competing principals. The
set of equilibrium allocations is very large and we identify it using
deviator-reporting mechanisms (DRMs) on the path and single transfer
schedules off the path.

We design a lab experiment implementing DRMs, which importantly characterize equilibria under model primitives and offer a simpler messaging system than in canonical mechanisms. We
observe that implemented outcomes are efficient more often than random. A
majority of the time, agents do tell the truth on the identity of a
deviating principal, despite potential gains from tacit collusion on false
reports. As play progresses, agents learn to play with their counterparty
agent, with the average predicted probability of collusion on false reports
across groups increasing from about 9\% at the beginning of the experiment
to just under 20\% by the end. However, group heterogeneity is significant, with predictions for some groups rising to 40\% by the end. Our paper complements a wide body of experimental literature on dishonesty, and is one of the first, to our knowledge, to provide empirical evidence on play in competing mechanism games. Our results hint at the limits of the equilibrium characterization based on truthful reporting in competing mechanisms.

\newpage

\section{Appendix}

\subsection{Appendix A: Proof of Proposition \protect\ref%
{prop:dev_to_transfer_schedules}\label{sec:proof:prop_dev_to_schedules}}

Fix an equilibrium $(\hat{\lambda},\hat{x},\hat{\sigma})$ where $\hat{\lambda%
}^{m}=\{\hat{\lambda}_{n}^{m}\}_{n\in \mathcal{N}}$ is given by, for all $%
n\in \mathcal{N}$
\begin{multline*}
\hat{\lambda}_{n}^{m}\left( k_{1}^{m},\ldots ,k_{N}^{m}\right) = \\
\left\{
\begin{array}{cc}
\hat{t}_{n}^{m,k} & \text{if }\exists k\neq m\text{ such that }\left\vert
\left\{ k_{n}^{m}\in \mathcal{M}:k_{n}^{m}=k,\text{ }n\in \mathcal{N}%
\right\} \right\vert >\frac{N}{2}; \\
&  \\
t_{n}^{\circ } &
\begin{array}{c}
\text{if }\exists k,k^{\prime }\neq m\text{ such that (i) }k\neq k^{\prime }%
\text{ and} \\
\text{(ii) }\left\vert \left\{ k_{n}^{m}\in \overline{\mathcal{M}}%
:k_{n}^{m}=k,\text{ }n\in \mathcal{N}\right\} \right\vert =\left\vert
\left\{ k_{n}^{m}\in \overline{\mathcal{M}}:k_{n}^{m}=k^{\prime },\text{ }%
n\in \mathcal{N}\right\} \right\vert =\frac{N}{2}%
\end{array}%
; \\
&  \\
\hat{t}_{n}^{m} & \text{otherwise.}%
\end{array}%
\right.
\end{multline*}%
Suppose that principal $m$ unilaterally deviates to a profile of transfer
schedules $t^{m}=\{t_{n}^{m}\}_{n\in \mathcal{N}}\in T^{m}$. We show that in
a truthful $\Lambda $ equilibrium, this is strategically equivalent to
unilaterally deviating to a DRM $\lambda ^{m}=\{\lambda _{n}^{m}\}_{n\in
\mathcal{N}}\in \Lambda ^{m}$ such that for all $n\in \mathcal{N}$,%
\begin{multline}
\lambda _{n}^{m}\left( k_{1}^{m},\ldots ,k_{N}^{m}\right) =  \label{DRM2} \\
\left\{
\begin{array}{cc}
t_{n}^{m,k} & \text{if }\exists k\neq m\text{ such that }\left\vert \left\{
k_{n}^{m}\in \mathcal{M}:k_{n}^{m}=k,\text{ }n\in \mathcal{N}\right\}
\right\vert >\frac{N}{2}, \\
&  \\
t_{n}^{\circ } &
\begin{array}{c}
\text{if }\exists k,k^{\prime }\neq m\text{ such that (i) }k\neq k^{\prime }%
\text{ and} \\
\text{(ii) }\left\vert \left\{ k_{n}^{m}\in \overline{\mathcal{M}}%
:k_{n}^{m}=k,\text{ }n\in \mathcal{N}\right\} \right\vert =\left\vert
\left\{ k_{n}^{m}\in \overline{\mathcal{M}}:k_{n}^{m}=k^{\prime },\text{ }%
n\in \mathcal{N}\right\} \right\vert =\frac{N}{2}%
\end{array}%
, \\
&  \\
t_{n}^{m} & \text{otherwise.}%
\end{array}%
\right.
\end{multline}%
The reason is that all agents report $m$ to principal $m$ since he is the
only deviator, and as a result, $t^{m}\in T^{m}$ is the assigned profile of
transfer schedules. Subsequently, in a truthful $\Lambda $ equilibrium upon
principal $m$'s deviation to $t^{m}\in T^{m}$, each agent chooses her
messages to the other principals and her action that are the same as ones
she chooses in a truthful $\Lambda $ equilibrium upon $m$'s deviation to $%
\lambda ^{m}\in \Lambda ^{m}$. Then, principal $m$'s payoff upon deviation
to $\lambda ^{m}\in \Lambda ^{m}$ is preserved by that upon deviation to $%
t^{m}\in T^{m}$. Since the deviation to $\lambda ^{m}\in \Lambda ^{m}$ is
not profitable for principal $m$, the deviation to $t^{m}\in T^{m}$ is not
profitable as well.

On the other hand, suppose that principal $m$ deviates to a DRM $\lambda
^{m}\in \Lambda ^{m}$ that satisfies (\ref{DRM2}). In a truthful $\Lambda $
equilibrium, this is strategically equivalent to deviating to a profile of
transfer schedules $t^{m}\in T^{m}.$ The reason is that in a truthful $%
\Lambda $ equilibrium upon $m$'s deviation to $\lambda ^{m}\in \Lambda ^{m}$%
, all agents report $m$ to principal $m$ and as a result, $t^{m}\in T^{m}$
is assigned. Subsequently, in a truthful $\Lambda $ equilibrium upon $m$'s
deviation to $\lambda ^{m}\in \Lambda ^{m}$, in addition to reporting $m$ to
principal $m,$ each agent chooses her messages to the other principals and
her action that are the same as ones she choose in a truthful $\Lambda $
equilibrium upon $m$'s deviation to $t^{m}\in T^{m}$. Then principal $m$'s
payoff upon deviation to $t^{m}\in T^{m}$ is preserved by that upon
deviation to $\lambda ^{m}\in T^{m}$. Because the deviation to $t^{m}\in
T^{m}$ is not profitable for principal $m,$ the deviation to $\lambda
^{m}\in \Lambda ^{m}$ is not profitable as well. $\blacksquare $

\subsection{Appendix B: Proof of Theorem \protect\ref%
{thm:equilibrium_characterization}\label{sec:pf:thm:eq_charzn}}

We first prove the \textquotedblleft only if\textquotedblright\ part. For
that, fix a truthful $\Lambda $ equilibrium $(\hat{\lambda},\hat{x},\hat{%
\sigma})$. We aim to show that $z^{(\hat{\lambda},\hat{x},\hat{\sigma})}\in
Z^{\ast }$. Because of Proposition \ref{prop:dev_to_transfer_schedules}, we
only need to consider principal $m$'s unilateral deviation to a profile of
transfer schedules, $t^{m}=\{t_{n}^{m}\}_{n\in \mathcal{N}}\in T^{m}$. Such
a deviation is equivalent to the deviation to a DRM that always assigns $%
t_{n}^{m}$ to agent $n$ except for the case where a half of agents report $k$
and the other half $k^{\prime }$ such that $k^{\prime }\neq k$. For
notational ease, we denote such a DRM by $t^{m}$

Let principal $m$'s equilibrium DRM $\hat{\lambda}^{m}=\{\hat{\lambda}%
_{n}^{m}\}_{n\in \mathcal{N}}$ have the following structure: For all $n\in
\mathcal{N}$,
\begin{multline*}
\hat{\lambda}_{n}^{m}\left( k_{1}^{m},\ldots ,k_{N}^{m}\right) = \\
\left\{
\begin{array}{cc}
\hat{t}_{n}^{m,k} & \text{if }\exists k\neq m\text{ such that }\left\vert
\left\{ k_{n}^{m}\in \mathcal{M}:k_{n}^{m}=k,\text{ }n\in \mathcal{N}%
\right\} \right\vert >\frac{N}{2}; \\
&  \\
t_{n}^{\circ } &
\begin{array}{c}
\text{if }\exists k,k^{\prime }\neq m\text{ such that (i) }k\neq k^{\prime }%
\text{ and} \\
\text{(ii) }\left\vert \left\{ k_{n}^{m}\in \overline{\mathcal{M}}%
:k_{n}^{m}=k,\text{ }n\in \mathcal{N}\right\} \right\vert =\left\vert
\left\{ k_{n}^{m}\in \overline{\mathcal{M}}:k_{n}^{m}=k^{\prime },\text{ }%
n\in \mathcal{N}\right\} \right\vert =\frac{N}{2}%
\end{array}%
; \\
&  \\
\hat{t}_{n}^{m} & \text{otherwise.}%
\end{array}%
\right.
\end{multline*}%
We have $\hat{t}^{m}=\{\hat{t}_{n}^{m}\}_{n\in \mathcal{N}}=\{\hat{\lambda}%
_{n}^{m}(\hat{x}^{m}(\hat{\lambda}))\}_{n\in \mathcal{N}}$ as the profile of
equilibrium transfer schedules. Let $\hat{t}=\{\hat{t}^{m}\}_{m\in \mathcal{M%
}}$. Given $\hat{s}=\{\hat{s}_{n}\}_{n\in \mathcal{N}}=\{\hat{\sigma}_{n}\{%
\hat{\lambda},\hat{x}_{n}(\hat{\lambda}),\hat{\lambda}(\hat{x}(\hat{\lambda}%
))\}_{n\in \mathcal{M}},$ let $\hat{d}_{n}^{m}=\hat{t}_{n}^{m}\left( \hat{s}%
_{n}\right) $ for all $n\in \mathcal{N}$ and all $m\in \mathcal{M}$ and $%
\hat{d}=\{\hat{d}_{n}^{m}\}_{m\in \mathcal{M},n\in \mathcal{N}}$. Then, the
equilibrium allocation is $z^{(\hat{\lambda},\hat{x},\hat{\sigma})}=(\hat{s},%
\hat{d}).$ Because $(\hat{\lambda},\hat{x},\hat{\sigma})$ is a $\Lambda $
equilibrium, $\hat{s}\in \hat{S}\left( \hat{t}\right) $ with $\hat{d}%
_{n}^{m}=\hat{t}_{n}^{m}\left( \hat{s}_{n}\right) .$ Therefore, qualifier
(i) in $Z^{\ast }$ is satisfied.

Note that $\hat{t}_{n}^{-m,m}=\{\hat{t}_{n}^{j,m}\}_{j\in \mathcal{M}%
\diagdown \{m\mathcal{\}}}$ is the profile of transfer schedules that the
other principals' DRMs $\hat{\lambda}^{-m}$ assign in a truthful $\Lambda $
equilibrium upon principal $m$'s unilateral deviation. Let $\hat{\sigma}%
_{n}[(t^{m},\hat{\lambda}^{-m}),\hat{x}_{n}(t^{m},\hat{\lambda}^{-m}),(t^{m},%
\hat{\lambda}^{-m}(\hat{x}^{-m}(t^{m},\hat{\lambda}^{-m}))]$ be agent $n$'s
action choice when principal $m$ unilaterally deviates to a DRM that always
assign $t^{m}$ except for the case where a half of agents report $k$ and the
other half report $k^{\prime }$ such that $k\neq k^{\prime }.$ Let
\begin{multline*}
\hat{\sigma}[(t^{m},\hat{\lambda}^{-m}),\hat{x}(t^{m},\hat{\lambda}%
^{-m}),(t^{m},\hat{\lambda}^{-m}(\hat{x}^{-m}(t^{m},\hat{\lambda}^{-m}))]= \\
\left\{ \hat{\sigma}_{n}[(t^{m},\hat{\lambda}^{-m}),\hat{x}_{n}(t^{m},\hat{%
\lambda}^{-m}),(t^{m},\hat{\lambda}^{-m}(\hat{x}^{-m}(t^{m},\hat{\lambda}%
^{-m}))]\right\} _{n\in \mathcal{N}}.
\end{multline*}%
Then, we have the following relations:%
\begin{eqnarray}
&&G^{m}(\hat{s})-\sum_{n\in \mathcal{N}}\hat{d}_{n}^{m}  \label{thm1_1} \\
&\geq &\max_{t^{m}\in T^{m}}\left\{
\begin{array}{c}
\ G^{m}[\hat{\sigma}[(t^{m},\hat{\lambda}^{-m}),\hat{x}(t^{m},\hat{\lambda}%
^{-m}),(t^{m},\hat{\lambda}^{-m}(\hat{x}^{-m}(t^{m},\hat{\lambda}^{-m}))]]
\\
-\sum_{n\in \mathcal{N}}t_{n}^{m}[\hat{\sigma}[(t^{m},\hat{\lambda}^{-m}),%
\hat{x}_{n}(t^{m},\hat{\lambda}^{-m}),(t^{m},\hat{\lambda}^{-m}(\hat{x}%
^{-m}(t^{m},\hat{\lambda}^{-m}))]]%
\end{array}%
\right\}  \notag \\
&\geq &\max_{t^{m}\in T^{m}}\left\{ \min_{s(\hat{t}^{-m,m},t^{m})\in \hat{S}%
(t^{-m,m},t^{m})}G^{m}(s\left( t^{m},\hat{t}^{-m,m}\right) )-\sum_{n\in
\mathcal{N}}t_{n}^{m}(s_{n}\left( t^{m},\hat{t}^{-m,m}\right) )\right\}
\notag \\
&\geq &\min_{t^{-m}\in T^{-m}}\max_{t^{m}\in T^{m}}\left\{
\min_{s(t^{m},t^{-m})\in \hat{S}(t^{m},t^{-m})}G^{m}(s\left(
t^{m},t^{-m}\right) )-\sum_{n\in \mathcal{N}}t_{n}^{m}(s_{n}\left(
t^{m},t^{-m}\right) )\right\} =\underline{V}^{m}  \notag
\end{eqnarray}%
The first inequality relation in (\ref{thm1_1}) holds because of Proposition %
\ref{prop:dev_to_transfer_schedules}. The second inequality relation holds
because
\begin{eqnarray*}
\hat{\lambda}^{-m}(\hat{x}^{-m}(t^{m},\hat{\lambda}^{-m})) &=&\hat{t}^{-m,m}%
\text{ and} \\
\hat{\sigma}[(t^{m},\hat{\lambda}^{-m}),\hat{x}(t^{m},\hat{\lambda}%
^{-m}),(t^{m},\hat{\lambda}^{-m}(\hat{x}^{-m}(t^{m},\hat{\lambda}^{-m})))]
&\in &\hat{S}(t^{-m,m},t^{m}).
\end{eqnarray*}%
The second inequality relation holds because $\hat{t}^{-m,m}\in T^{-m}.$ (%
\ref{thm1_1}) implies (ii) in $Z^{\ast }$ is satisfied. Therefore, $z^{(\hat{%
\lambda},\hat{x},\hat{\sigma})}\in Z^{\ast }$.

Now we prove the \textquotedblleft if\textquotedblright\ part. For that, fix
$(\hat{s},\hat{d})\in Z^{\ast }$. We aim to show that there exists a
truthful equilibrium such that $z^{(\hat{\lambda},\hat{x},\hat{\sigma})}=(%
\hat{s},\hat{d})$. Because $(\hat{s},\hat{d})\in Z^{\ast }$, there is a
profile of transfer schedules $\hat{t}$ such that $\hat{s}\in \hat{S}(t)\ $%
and$\ \hat{d}_{n}^{m}=\hat{t}_{n}^{m}(\hat{s}_{n})$. For all $k\in \mathcal{M%
}$, we pick $\tilde{t}^{-m,m}$ such that%
\begin{equation}
\tilde{t}^{-m,m}\in \arg \min_{t^{-m}\in T^{-m}}\max_{t^{m}\in T^{m}}\left\{
\min_{s(t^{m},t^{-m})\in \hat{S}(t^{m},t^{-m})}G^{m}(s(t^{m},t^{-m}))-%
\sum_{n\in \mathcal{N}}t_{n}^{m}(s_{n}(t^{m},t^{-m}))\right\}
\label{thm1_1A}
\end{equation}%
In order to support $(\hat{s},\hat{d})$ as an equilibrium allocation, we let
principal $m$ offer a DRM $\hat{\lambda}^{m}$ such that%
\begin{multline}
\hat{\lambda}_{n}^{m}\left( k_{1}^{m},\ldots ,k_{N}^{m}\right) =
\label{thm1_1B} \\
\left\{
\begin{array}{cc}
\tilde{t}_{n}^{m,k} & \text{if }\exists k\neq m\text{ such that }\left\vert
\left\{ k_{n}^{m}\in \mathcal{M}:k_{n}^{m}=k,\text{ }n\in \mathcal{N}%
\right\} \right\vert >\frac{N}{2}, \\
&  \\
t_{n}^{\circ } &
\begin{array}{c}
\text{if }\exists k,k^{\prime }\neq m\text{ such that (i) }k\neq k^{\prime }%
\text{ and} \\
\text{(ii) }\left\vert \left\{ k_{n}^{m}\in \overline{\mathcal{M}}%
:k_{n}^{m}=k,\text{ }n\in \mathcal{N}\right\} \right\vert =\left\vert
\left\{ k_{n}^{m}\in \overline{\mathcal{M}}:k_{n}^{m}=k^{\prime },\text{ }%
n\in \mathcal{N}\right\} \right\vert =\frac{N}{2}%
\end{array}%
, \\
&  \\
\hat{t}_{n}^{m} & \text{otherwise.}%
\end{array}%
\right.
\end{multline}%
If no principal deviates, the agent's communication and action strategy
follow:%
\begin{align}
\hat{x}_{n}^{m}(\hat{\lambda})& =0\text{ for all }\left( m,n\right) \in
\mathcal{M}\mathbb{\times }\mathcal{N}  \label{thm1_2} \\
\hat{\sigma}_{n}[\hat{\lambda},\hat{x}_{n}(\hat{\lambda}),\hat{\lambda}(\hat{%
x}(\hat{\lambda}))]& =\hat{s}_{n}\text{ for all }n\in \mathcal{N}
\label{thm1_3}
\end{align}%
(\ref{thm1_2}) and (\ref{thm1_3}) imply that
\begin{eqnarray}
\hat{\lambda}_{n}^{m}(\hat{x}^{m}(\hat{\lambda})) &=&\hat{t}_{n}^{m}
\label{thm1_4} \\
\hat{\lambda}_{n}^{m}(\hat{x}^{m}(\hat{\lambda}))[\hat{\sigma}_{n}(\hat{%
\lambda},\hat{x}_{n}(\hat{\lambda}),\hat{\lambda}(\hat{x}(\hat{\lambda}))]]
&=&\hat{t}_{n}^{m}\left( \hat{s}\right) =\hat{d}_{n}^{m}  \label{thm1_5}
\end{eqnarray}%
Furthermore, $\hat{s}\in \hat{S}(\hat{t})$ and hence $\hat{\sigma}=\{\hat{%
\sigma}_{n}\}_{n\in \mathcal{N}}$ satisfying (\ref{thm1_3}) characterizes
each agent $n$'s optimal action choice given $\hat{\lambda}_{n}(\hat{x}(\hat{%
\lambda}))=\hat{t}_{n}$ as the profile of transfer schedules that are
assigned to her. Note that principal $m$'s DRM offers $t_{n}^{\circ }$ to
each agent $n$ if the half of agents send $k$ and the other half $k^{\prime
} $ such that $k^{\prime }\neq k.$ This makes it weakly dominated for an
agent to send a false message to principal $m$ when every other agent sends
a truthful report. Therefore, $\hat{x}=\{\hat{x}^{m}\}_{m\in \mathcal{M}}$
satisfying (\ref{thm1_2}) shows that each agent's message is truthful on the
equilibrium path and it is a best response to every other agent's truthful
messages given $\hat{\lambda}$. Therefore, $\left( \hat{x},\hat{\sigma}%
\right) $ characterizes a profile of equilibrium communication and action
strategies when no principals deviates. (\ref{thm1_3}) and (\ref{thm1_5})
imply that $z^{\left( \hat{\lambda},\hat{x},\hat{\sigma}\right) }=(\hat{s},%
\hat{d})\in Z^{\ast }$.

Because of Proposition \ref{prop:dev_to_transfer_schedules}, we only need to
consider principal $m$'s unilateral deviation to a profile of transfer
schedules $t^{m}=\{t_{n}^{m}\}_{n\in \mathcal{N}}\in T^{m}$. Such a
deviation is equivalent to the deviation to a DRM that always assigns $%
t_{n}^{m}$ to agent $n$ except for the case where a half of agents report $k$
and the other half $k^{\prime }$ such that $k^{\prime }\neq k$. For
notational ease, we denote such a DRM by $t^{m}$. Upon such a deviation,
agents' communication and action strategies follow:%
\begin{equation}
\hat{x}_{n}^{k}(t^{m},\hat{\lambda}^{-m})=m\text{ for all }\left( k,n\right)
\in \mathcal{M}\mathbb{\times }\mathcal{N}  \label{thm1_6}
\end{equation}%
and
\begin{gather}
\{\hat{\sigma}_{n}[(t^{m},\hat{\lambda}^{-m}),\hat{x}_{n}(t^{m},\hat{\lambda}%
^{-m}),(t^{m},\hat{\lambda}^{-m}(\hat{x}^{-m}(t^{m},\hat{\lambda}%
^{-m}))]\}_{n\in \mathcal{N}}\in  \label{thm1_7} \\
\arg \min_{s(t^{m},\tilde{t}^{-m,m})\in \hat{S}(t^{m},\tilde{t}^{-m,m})}%
\left[ G^{m}(s(t^{m},\tilde{t}^{-m,m}))-\sum_{n\in \mathcal{N}%
}t_{n}^{m}(s_{n}(t^{m},\tilde{t}^{-m,m}))\right]  \notag
\end{gather}%
(\ref{thm1_6}) shows that each agent $n$ reports the true message to every
principal (i.e., $m$ is the identity of the deviating principal) and it is a
best-response report for every agent when every other agent reports the true
message. Because of (\ref{thm1_6}), the profile of non-deviating principals'
transfer schedules become $\hat{\lambda}_{n}^{-m}(\hat{x}^{-m}(t^{m},\hat{%
\lambda}^{-m}))=\tilde{t}_{n}^{-m,m}$ given their equilibrium DRMs and (\ref%
{thm1_7}) implies that each agent $n$'s action choice
\begin{equation*}
\hat{\sigma}_{n}[(t^{m},\hat{\lambda}^{-m}),\hat{x}_{n}(t^{m},\hat{\lambda}%
^{-m}),(t^{m},\hat{\lambda}^{-m}(\hat{x}^{-m}(t^{m},\hat{\lambda}^{-m}))]
\end{equation*}%
is optimal. Therefore, $\left( \hat{x},\hat{\sigma}\right) $ satisfying (\ref%
{thm1_6}) and (\ref{thm1_7}), characterize a truthful $\Lambda $ equilibrium
upon principal $m$'s unilateral deviation to a profile of transfer schedules
$t^{m}.$ Given such a truthful $\Lambda $ equilibrium, principal $m$'s
payoff upon his unilateral deviation to $t^{m}$ is
\begin{align}
& V^{m}((t^{m},\hat{\lambda}^{-m}),\hat{x},\hat{\sigma})  \label{thm1_8} \\
& =G^{m}[\hat{\sigma}[(t^{m},\hat{\lambda}^{-m}),\hat{x}(t^{m},\hat{\lambda}%
^{-m}),(t^{m},\hat{\lambda}^{-m}(\hat{x}^{-m}(t^{m},\hat{\lambda}^{-m})))]]
\notag \\
& -\sum_{n\in \mathcal{N}}t_{n}^{m}[\hat{\sigma}[(t^{m},\hat{\lambda}^{-m}),%
\hat{x}(t^{m},\hat{\lambda}^{-m}),(t^{m},\hat{\lambda}^{-m}(\hat{x}%
^{-m}(t^{m},\hat{\lambda}^{-m})))]]  \notag \\
& =\min_{s(t^{m},\tilde{t}^{-m,m})\in \hat{S}(t^{m},\tilde{t}^{-m,m})}\left[
G^{m}(s\left( t^{m},\tilde{t}^{-m,m}\right) )-\sum_{n\in \mathcal{N}%
}t_{n}^{m}(s_{n}\left( t^{m},\tilde{t}^{-m,m}\right) )\right]  \notag \\
& \leq \min_{t^{-m}\in T^{-m}}\max_{t^{m}\in T^{m}}\left\{
\min_{s(t^{-m},t^{m})\in \hat{S}(t^{-m},t^{m})}\left[
G^{m}(s(t^{m},t^{-m}))-\sum_{n\in \mathcal{N}}t_{n}^{m}(s_{n}(t^{m},t^{-m}))%
\right] \right\} =\underline{V}^{m}  \notag
\end{align}%
The second equality relation holds because of (\ref{thm1_7}). The inequality
relation holds because $t^{m}$ is one of possible profiles of transfer
schedules in $T^{m}$, given the definition of $\tilde{t}^{-m,m}$ in (\ref%
{thm1_1A}).

On the other hand, because $z^{(\hat{\lambda},\hat{x},\hat{\sigma})}=(\hat{s}%
,\hat{d})\in Z^{\ast }$, we have that
\begin{eqnarray}
&&V^{m}(\hat{\lambda},\hat{x},\hat{\sigma})  \label{thm1_9} \\
&=&G^{m}[\hat{\sigma}(\hat{\lambda},\hat{x}(\hat{\lambda}),\hat{\lambda}(%
\hat{x}(\hat{\lambda})))]-\sum_{n\in \mathcal{N}}t_{n}^{m}[\hat{\sigma}_{n}(%
\hat{\lambda},\hat{x}_{n}(\hat{\lambda}),\hat{\lambda}(\hat{x}(\hat{\lambda}%
)))]  \notag \\
&\geq &\underline{V}^{m}  \notag
\end{eqnarray}%
Combining (\ref{thm1_8}) and (\ref{thm1_9}) yields that for all $t^{m}\in
T^{m}$,%
\begin{equation}
V^{m}(\hat{\lambda},\hat{x},\hat{\sigma})\geq V^{m}((t^{m},\hat{\lambda}%
^{-m}),\hat{x},\hat{\sigma}).  \label{thm1_10}
\end{equation}%
Because we can assign an arbitrary continuation equilibrium upon multiple
principals' deviations, (\ref{thm1_2}), (\ref{thm1_3}), (\ref{thm1_6}), (\ref%
{thm1_7}), and (\ref{thm1_10}) shows that there is a truthful equilibrium $(%
\hat{\lambda},\hat{x},\hat{\sigma})$ with $z^{(\hat{\lambda},\hat{x},\hat{%
\sigma})}=(\hat{s},\hat{d})$. $\blacksquare $

\subsection{Appendix C: Proof of Corollary \label{sec:pf:corollary}}

Pick any equilibrium $(\hat{t},\hat{\sigma})\in \mathcal{E}^{G}$. The
equilibrium allocation is then
\begin{equation*}
(\hat{s},\hat{d})\equiv z^{(\hat{t},\hat{\sigma})}=[\left\{ \hat{\sigma}%
_{n}\left( \hat{t}\right) \right\} _{n\in \mathcal{N}}\text{, }\left\{ \hat{t%
}_{n}^{m}\left( \hat{\sigma}_{n}\left( \hat{t}\right) \right) \right\}
_{m\in \mathcal{M}\text{, }n\in \mathcal{N}}]\in S\times
%TCIMACRO{\U{211d} }%
%BeginExpansion
\mathbb{R}
%EndExpansion
_{+}^{M\times N}
\end{equation*}%
Condition 1 in Definition \ref{def:G-equilibrium} implies that $\hat{\sigma}%
\left( \hat{t}\right) =\left\{ \hat{\sigma}_{n}\left( \hat{t}\right)
\right\} _{n\in \mathcal{N}}\in \hat{S}\left( \hat{t}\right) $ and hence
qualifier (i) in $Z^{\ast }$ is satisfied. We also have the following
inequality relations:
\begin{align}
G^{m}(\hat{s})-\sum_{n\in \mathcal{N}}\hat{d}_{n}^{m}& \geq \max_{t^{m}\in
T^{m}}\left\{ G^{m}(\hat{\sigma}(t^{m},\hat{t}^{-m}))-\sum_{n\in \mathcal{N}%
}t_{n}^{m}(\hat{\sigma}_{n}(t^{m},\hat{t}^{-m}))\right\}  \label{Cor1_1} \\
& \geq \max_{t^{m}\in T^{m}}\left\{ \min_{s\in \hat{S}\left( t^{m},\hat{t}%
^{-m}\right) }\left[ G^{m}(s)-\sum_{n\in \mathcal{N}}t_{n}^{m}(s)\right]
\right\}  \notag \\
& \geq \min_{t^{-m}\in T^{-m}}\max_{t^{m}\in T^{m}}\left\{ \min_{s\in \hat{S}%
\left( t^{m},\hat{t}^{-m}\right) }\left[ G^{m}(s)-\sum_{n\in \mathcal{N}%
}t_{n}^{m}(s)\right] \right\} =\underline{V}^{m}.  \notag
\end{align}%
Note that the expression on the left-hand side on the first line is
principal $m$'s equilibrium payoff. The first inequality is due to Condition
2 in Definition \ref{def:G-equilibrium}. The second inequality holds because
$\hat{\sigma}(t^{m},\hat{t}^{-m})\in \hat{S}\left( t^{m},\hat{t}^{-m}\right)
.$ The third inequality holds because $\hat{t}^{-m}\in $ $T^{-m}.$ (\ref%
{Cor1_1}) implies that qualifier (ii) in $Z^{\ast }$ is satisfied. Because
both qualifiers (i) and (ii) in $Z^{\ast }$ are satisfied, $z^{(\hat{t},\hat{%
\sigma})}\in Z^{\ast }$. $\blacksquare $

\subsection{Appendix D: Proof of Proposition \protect\ref{prop:2by2}\label%
{sec:pf:prop2by2}}

First, notice that whenever $G^{m}(\sigma
(t^{m},t^{-m}))-\sum_{i=1}^{2}t_{i}^{m}(\sigma _{i}(t^{m},t^{-m})\leq
\underline{G}^{m}$, principal $m$ has a profitable deviation to $\tilde{t}%
^{m}=t^{\circ }$, the zero transfer schedule, since they can ensure at least
the minimum gross payoffs from the game when the sum of their transfers
equals zero. Therefore, for an equilibrium to be supported it must be that
the DRM of each principal $m$ features transfer schedules such that%
\begin{equation}
G^{m}(\sigma (t^{m},t^{-m}))-\sum_{i=1}^{2}t_{i}^{m}(\sigma
_{i}(t^{m},t^{-m})\geq \underline{G}^{m}  \label{devcond2}
\end{equation}%
\noindent when outcome $\sigma (t^{m},t^{-m})$ is implemented. Construct a
DRM such that%
\begin{equation}
\underline{G}^{m}\geq G^{m}(\sigma (\tilde{t}^{m},t^{-m,m}))-\sum_{i=1}^{2}%
\tilde{t}_{i}^{m}(\sigma _{i}(\tilde{t}^{m},t^{-m,m}))  \label{devcond1}
\end{equation}%
\noindent holds. Then under this DRM, (\ref{devcond2}) and (\ref{devcond1})
imply

\begin{equation*}
G^{m}(\sigma (t^{m},t^{-m}))-\sum_{i=1}^{2}t_{i}^{m}(\sigma
_{i}(t^{m},t^{-m})\geq G^{m}(\sigma (\tilde{t}^{m},t^{-m,m}))-\sum_{i=1}^{2}%
\tilde{t}_{i}^{m}(\sigma _{i}(t^{m},t^{-m})).
\end{equation*}

\noindent Let principal 2 be the arbitrary deviator and construct such a DRM
\begin{equation*}
\lambda _{n}^{1}\left( k_{1}^{1},k_{2}^{1}\right) =\left\{
\begin{array}{cc}
t_{n}^{1,2} & \text{if }k_{1}^{1}=k_{2}^{1}=2 \\
&  \\
t_{n}^{\circ } &
\begin{array}{c}
\text{if }k_{1}^{1}\neq k_{2}^{1}%
\end{array}%
; \\
&  \\
t_{n}^{1} & \text{otherwise.}%
\end{array}%
\right.
\end{equation*}%
where ${t}_{n}^{1,2}$ offers only positive amounts for outcomes with $G^{2}=%
\underline{G}^{2}$ to punish the deviating principal. Without loss of
generality, let $\underline{G}^{2}=y_{1}$, then ${t}%
^{1,2}=(t_{A}^{1,2},t_{B}^{1,2})=((t_{A}^{1,2}(U),t_{A}^{1,2}(D)),(t_{B}^{1,2}(L),t_{B}^{1,2}(R)))
$ where $t_{A}^{1,2}(U)>0,t_{B}^{1,2}(L)>0,$ and $%
t_{A}^{1,2}(D)=t_{B}^{1,2}(R)=0$. If principal 2 deviates, they must
consider the transfer schedule $t^{1,2}$ when bidding for an agent's action.
To implement an action other than $(U,L)$ upon deviating, principal 2 must
offer $\tilde{t}_{A}^{2}(D)\geq t_{A}^{1,2}(U)$ to implement $(D,L)$, $%
\tilde{t}_{B}^{2}(R)\geq t_{B}^{1,2}(L)$ to implement $(U,R)$, or outbid for
both agents to implement the opposite diagonal $(D,R)$ at a cost of $\tilde{t%
}_{A}^{2}(D)+\tilde{t}_{B}^{2}(R)$.

To prevent these offers from being accepted and profitable, principal 1's
DRM offers $t_{A}^{1,2}(U)\geq y_{3}$, $t_{B}^{1,2}(L)\geq y_{2}$, and $%
t_{A}^{1,2}(U)+t_{B}^{1,2}(L)\geq y_{4}$. Hence, to implement another
outcome when facing principal 1's DRM with such a punishing transfer
schedule, principal 2 must pay either $\tilde{t}_{A}^{2}(D)>t_{A}^{1,2}(U)%
\geq y_{3}$, $\tilde{t}_{B}^{2}(R)>t_{B}^{1,2}(L)\geq y_{2}$, or $\tilde{t}%
_{A}^{2}(D)+\tilde{t}_{B}^{2}(R)>y_{4}$. However, all of these transfer
amounts are in excess of the gross payoffs that are received when moving to
that outcome, and are therefore not profitable. In a similar fashion, again
without loss of generality considering $\underline{G}^{1}=x_{1}$, principal
2 constructs a DRM such that principal 1 must pay more than their gross
payoffs to implement a different outcome. That is, principal 2's DRM offers $%
t_{A}^{2,1}(U)\geq x_{3}$, $t_{B}^{2,1}(L)\geq x_{2}$, and $%
t_{A}^{2,1}(U)+t_{B}^{2,1}(L)\geq x_{4}$.

Under these schedules, a deviating principal $m$ can get at most their
minimum gross payoff. Therefore, a DRM with $t^{m,-m}$ constructed as above
and with any transfer schedules $t^{1}=(t_{A}^{1},t_{B}^{1})$ and $%
t^{2}=(t_{A}^{2},t_{B}^{2})$ such that (\ref{devcond2}) is satisfied will
implement outcome $\sigma (t^{1},t^{2})$ in equilibrium, which can be any
outcome if transfers are sufficiently low to maintain (\ref{devcond2}). $%
\blacksquare $

\subsection{Appendix E: $\Gamma $ equilibrium\label{sec:CMGPTA_Gamma}}

The set of equilibrium allocations is subject to the set of mechanisms
allowed in a game. Such a dependence generates two potential problems of an
equilibrium allocation. An equilibrium allocation derived with a restricted
set of mechanisms may disappear. There can be new equilibrium allocations
that can be supported if a bigger set of mechanisms are allowed in a game.
In this section, we show that the set of truthful $\Lambda $ equilibrium
allocations, $Z^{\ast }$ is the set of equilibrium allocations that can be
supported with any set of complex mechanisms allowed in a game if the game
is \textquotedblleft regular\textquotedblright .

We allow each principal to send a message to himself. As shown below, this
feature makes it no loss of generality for agents to rely on a binary
message for communication with principals on the equilibrium path in the
CMGPTA (i.e., the use of the DRM on the path). Let $C_{0}^{m}$ be the set of
messages that principal $m$ can send to himself. Let $C_{n}^{m}$ be the set
of messages that agent $n$ can send to principal $m$, $C^{m}\equiv \Pi
_{n\in \mathcal{N}}C_{n}^{m}$ and $C_{n}\equiv \Pi _{m\in \mathcal{M}%
}C_{n}^{m}$. Principal $j$ offers a contract $\gamma
_{n}^{m}:C^{m}\rightarrow T_{n}^{m}$ to agent $m.$ Given a contract $\gamma
_{n}^{m}$ and a profile of messages $c^{m}=\{c_{n}^{m}\}_{n\in \mathcal{N}%
}\in C^{m}$, principal $m$'s transfer schedule for agent $n$ becomes $\gamma
_{n}^{m}\left( c^{m}\right) $. Let $c_{n}$ denote a profile of messages sent
by agent $n$ to all principals, i.e., $c^{n}=\{c_{n}^{m}\}_{m\in \mathcal{M}%
}\in C_{n}.$ Let $C\equiv \Pi _{m\in \mathcal{M}}C^{m}$

Let $\Gamma _{n}^{m}$ be the set of all contracts that principal $m$ can
offer to agent $n.$ Principal $m$'s mechanism $\gamma ^{m}$ is a profile of
contracts $\gamma ^{m}=\{\gamma _{n}^{m}\}_{n\in \mathcal{N}}\Gamma
^{m}\equiv \Pi _{n\in \mathcal{N}}\Gamma _{n}^{m}$. The nature of
communication that the message space $C^{m}$ permits can be quite general
such as mechanisms offered by other principals, transfer schedules that
would emerge as a result of communication with other principals, etc.

The timing of the game relative to $\Gamma $ is as follows.

\begin{enumerate}
\item Each principal $m$ announces a mechanism $\gamma ^{m}\in \Gamma ^{m}$.
agents observe all the mechanisms but principal $m$ observes only his
mechanism.

\item After observing the profile of mechanisms $\gamma =\{\gamma
^{m}\}_{m\in \mathcal{M}}$ announced by principals, each agent $n$ privately
sends a message $c_{n}^{m}$ to each principal $m$ who also sends a message $%
c_{0}^{m}$ to himself at the same time without observing the other
principals' mechanisms.

\item Given the transfer schedules, each agent $n$ chooses action $s_{n}.$

\item Payoffs are assigned as in GPTA. principals receive gross payoffs
according to $G(s)$ minus the sum of payments they make for the actions.
agents receive the sum of payments owed to them for $s_{n}$ plus some
action-specific payoff or loss $F_{n}(s_{n})$.
\end{enumerate}

A strategy for principal $m$ is simply some mechanism $\gamma ^{m}\in \Gamma
^{m}$. Agent $n$' communication strategy is a profile of functions $%
x_{n}=\{x_{n}^{m}\}_{m\in \mathcal{M}}$, where each $x_{n}^{m}$ is a
function from $\Gamma $ into $C_{n}^{m}$. Hence, $x_{n}^{m}(\gamma )\in
C_{n}^{m}$ is the message agent $n$ sends to principal $m$. Each principal $%
m $'s communication strategy is a function $x_{0}^{m}$ from $\Gamma _{m}$
into $C_{0}^{m}.$ Given $\gamma \in \Gamma $, let $x_{n}(\gamma )=\left\{
x_{n}^{m}(\gamma )\right\} _{m\in \mathcal{M}}$, $x_{\mathcal{N}}(\gamma
)=\left\{ x_{n}(\gamma )\right\} _{n\in \mathcal{N}}$, $x_{-0}^{m}\left(
\gamma \right) =\left\{ x_{n}^{m}(\gamma )\right\} _{n\in \mathcal{N}}$, $%
x^{m}\left( \gamma \right) =\left( x_{0}^{m}\left( \gamma ^{m}\right)
,x_{-0}^{m}(\gamma )\right) ,$ $x^{-m}\left( \gamma \right) =\left\{
x^{j}\left( \gamma \right) \right\} _{j\in \mathcal{M\diagdown }\left\{
m\right\} }$, $x_{-n}^{m}\left( \gamma \right) =\left\{ x_{i}^{m}\left(
\gamma \right) \right\} _{i\in \{0\}\cup \mathcal{N\diagdown }\left\{
n\right\} },$ and $x_{-n}\left( \gamma \right) =\left\{ x_{-n}^{m}\left(
\gamma \right) \right\} _{m\in \mathcal{M}}$.

Given $\gamma =\{\gamma ^{m}\}_{m\in \mathcal{M}},$ let $\gamma _{n}\left(
x(\gamma )\right) =\left\{ \gamma _{n}^{m}\left( x^{m}(\gamma )\right)
\right\} _{m\in \mathcal{M}}$, $\gamma \left( x(\gamma )\right) =\left\{
\gamma _{n}\left( x(\gamma )\right) \right\} _{n\in \mathcal{N}}$, $\gamma
_{n}^{-m}\left( x^{-m}\left( \gamma \right) \right) =\{\gamma _{n}^{j}\left(
x^{j}(\gamma )\right) \}_{k\in \mathcal{M\diagdown }\left\{ m\right\} }$.
Given $n\in \mathcal{N}$, $\gamma =\{\gamma ^{m}\}_{m\in \mathcal{M}}$, $%
x_{-n}\left( \gamma \right) $ and $c_{n}=\{c_{n}^{m}\}_{m\in \mathcal{M}},$
let $\gamma _{n}\left( c_{n},x_{-n}(\gamma )\right) =\left\{ \gamma
_{n}^{m}\left( c_{n}^{m},x_{-n}^{m}(\gamma )\right) \right\} _{m\in \mathcal{%
M}}$ denote a profile of transfer schedules for agent $n$ when she sends
message $c_{n}^{m}$ and principal $m$ and the other agents send a profile of
messages $x_{-n}^{m}\left( \gamma \right) $ to principal $m$ for all $m\in
\mathcal{M}$.

Each agent $n$'s action strategy is denoted by $\sigma _{n}:\Gamma \times
C_{n}\times T\rightarrow S_{n}$, and $\sigma _{n}(\gamma ,c_{n},t)$ is her
chosen action when $\gamma $ is a profile of mechanisms, $c_{n}$ is a
profile of messages she sends, and $t_{n}$ is a profile of transfer
schedules assigned to her. A profile of agents' action strategies is then $%
\sigma =(\sigma _{1},\dots ,\sigma _{N})$. For all $\gamma \in \Gamma $, all
$m\in \mathcal{M}$, and all $c_{0}^{m}\in C_{0}^{m}$, we sometimes simplify
the notation of agent $n$'s action as follows:%
\begin{equation*}
s_{n}\left( \sigma _{n},\gamma ,c_{0}^{m},x_{-0}^{m},x^{-m}\right) =\sigma
_{n}[\gamma ,x_{n}(\gamma ),(\gamma _{n}^{m}(c_{0}^{m},x_{-0}^{m}(\gamma
)),\gamma _{n}^{-m}(x^{-m}(\gamma )))],
\end{equation*}%
which is agent $n$'s action given her action strategy $\sigma _{n}$, when $%
\gamma $ is the profile of mechanisms and all players follow their
communication strategies $(x_{-0}^{m},x^{-m})$ except principal $m$ sending $%
c_{0}^{m}$ to himself. Let $s\left( \sigma ,\gamma
,c_{0}^{m},x_{-0}^{m},x^{-m}\right) =\left\{ s_{n}\left( \sigma _{n},\gamma
,c_{0}^{m},x_{-0}^{m},x^{-m}\right) \right\} _{n\in \mathcal{N}}$. For all $%
\gamma \in \Gamma $, all $m\in \mathcal{M}$, we also simplify the notation
of agent $n$'s action as follows:%
\begin{equation*}
s_{n}\left( \sigma _{n},\gamma ,x\right) =\sigma _{n}[\gamma ,x_{n}(\gamma
),\gamma \left( x\left( \gamma \right) \right) ],
\end{equation*}%
which is agent $n$'s action given her strategy $\sigma _{n}$, when $\gamma $
is the profile of mechanisms and all players follow their communication
strategies $x$. Let $s\left( \sigma ,\gamma ,x\right) =\left\{ s_{n}\left(
\sigma _{n},\gamma ,x\right) \right\} _{n\in \mathcal{N}}$

\begin{definition}
\label{def:gamma_eq}$\{\hat{\gamma},\hat{x},\hat{\sigma}\}$ is a $\Gamma $
equilibrium if it satisfies the following conditions:

\begin{enumerate}
\item for all $n\in \mathcal{N}$ and all $\left( \gamma ,c_{n},t\right) \in
\Gamma \times C_{n}\times T$:
\begin{equation*}
\hat{\sigma}_{n}\left( \gamma ,c_{n},t\right) \in \arg \max_{s_{n}\in
S_{n}}\left\{ F_{n}(s_{n})+\sum_{k\in \mathcal{M}}t_{n}^{k}(s_{n})\right\} ,
\end{equation*}

\item for all $n\in \mathcal{N}$ and all $\gamma \in \Gamma $:%
\begin{equation*}
\hat{x}_{n}(\gamma )\in \arg \max_{c_{n}=\left( c_{n}^{1},\ldots
c_{n}^{M}\right) \in C_{n}}\left\{
\begin{array}{c}
F_{n}\left[ \hat{\sigma}_{n}\left( \gamma ,c_{n},\gamma (c_{n},\hat{x}%
_{-n}(\gamma ))\right) \right] + \\
\sum_{m\in \mathcal{M}}\gamma _{n}^{m}(c_{n}^{m},\hat{x}_{-n}^{m}(\gamma ))%
\left[ \hat{\sigma}_{n}\left( \gamma ,c_{n},\gamma (c_{n},\hat{x}%
_{-n}(\gamma ))\right) \right]%
\end{array}%
\right\} ,
\end{equation*}

\item for all $m\in \mathcal{M}$, and all $\gamma ^{m}\in \Gamma ^{m}$:
\begin{equation*}
\hat{x}_{0}^{m}\left( \gamma ^{m}\right) \in \arg \max_{c_{0}^{m}\in
C_{0}^{m}}\left\{
\begin{array}{c}
G^{m}\left[ s\left( \hat{\sigma},\gamma ^{m},\hat{\gamma}%
^{-m},c_{0}^{m},x_{-0}^{m},x^{-m}\right) \right] - \\
\sum_{n\in \mathcal{N}}\gamma _{n}^{m}(c_{0}^{m},\hat{x}_{-0}^{m}(\gamma
^{m},\hat{\gamma}^{-m}))\left[ s_{n}\left( \hat{\sigma}_{n},\gamma ^{m},\hat{%
\gamma}^{-m},c_{0}^{m},\hat{x}_{-0}^{m},\hat{x}^{-m}\right) \right]%
\end{array}%
\right\} ,
\end{equation*}

\item for all $m\in \mathcal{M}$:%
\begin{equation*}
\hat{\gamma}^{m}\in \arg \max_{\gamma ^{m}\in \Gamma ^{m}}\left\{
\begin{array}{c}
G^{m}\left[ s\left( \hat{\sigma},\gamma ^{m},\hat{\gamma}^{-m},\hat{x}%
\right) \right] - \\
\sum_{n\in \mathcal{N}}\gamma _{n}^{m}(\hat{x}^{m}(\gamma ^{m},\hat{\gamma}%
^{-m}))\left[ s_{n}\left( \hat{\sigma}_{n},\gamma ^{m},\hat{\gamma}^{-m},%
\hat{x}\right) \right]%
\end{array}%
\right\} .
\end{equation*}
\end{enumerate}
\end{definition}

Note that principal $m$ sends a message to himself at the same time agents
send messages to him. Condition 3 shows that principal $m$ sends an optimal
message to himself, taking as given agents' communication strategies, the
other principals' mechanism strategies, and their strategies of
communicating with themselves. $\left( \hat{x},\hat{\sigma}\right) $
satisfying Conditions 1, 2, 3 in Definition \ref{def:gamma_eq} constitutes a
continuation equilibrium given every possible profile of mechanisms in $%
\Gamma $.

Let $\mathcal{E}^{\Gamma }$ be the set of all $\Gamma $ equilibria. For any $%
\left( \hat{\gamma},\hat{x},\hat{\sigma}\right) \in \mathcal{E}^{\Gamma }$,
the equilibrium allocation is
\begin{eqnarray*}
z^{\left( \hat{\gamma},\hat{x},\hat{\sigma}\right) } &\equiv &\left( \left[
\hat{\sigma}_{n}\left( \hat{\gamma},\hat{x}_{n}(\hat{\gamma}),\left[ \hat{%
\gamma}(\hat{x}(\hat{\gamma}))\right] \right) \right] _{n\in \mathcal{N}},%
\left[ \hat{\gamma}_{n}^{m}(\hat{x}^{m}(\hat{\gamma}))\left( \hat{\sigma}%
_{n}\left( \hat{\gamma},\hat{x}_{n}(\hat{\gamma}),\hat{\gamma}(\hat{x}(\hat{%
\gamma}))\right) \right) \right] _{n\in \mathcal{N}}\right) \\
&\in &S\times \mathbb{R}_{+}^{M\times N}
\end{eqnarray*}%
Let
\begin{equation*}
Z^{\Gamma }\equiv \left\{ z^{\left( \hat{\gamma},\hat{x},\hat{\sigma}\right)
}:\left( \hat{\gamma},\hat{x},\hat{\sigma}\right) \in \mathcal{E}^{\Gamma
}\right\}
\end{equation*}%
be the set of all $\Gamma $ equilibrium allocations.

Now we define the \textquotedblleft regular\textquotedblright\ property of
the game relative to $\Gamma $. For every $m\in \mathcal{M}$, let $\tilde{%
\gamma}^{-m}=\left\{ \tilde{\gamma}^{j}\right\} _{j\in \mathcal{M\diagdown }%
\left\{ m\right\} }$ be the profile of mechanisms that assign $\tilde{t}%
^{-m,m}$ regardless of messages that the principals receive, that is,
principal $k$'s mechanism $\tilde{\gamma}^{k}$ always assigns $\tilde{t}%
^{k,m}$ regardless of messages principal $k$ receives. $\tilde{\gamma}^{k}$
is strategically equivalent to $\tilde{t}^{k,m}.$

\begin{definition}
\label{def:regular}The game relative to $\Gamma $ is regular if

\begin{enumerate}
\item $\left\vert C_{n}^{m}\right\vert \geq \left\vert \overline{\mathcal{M}}%
\right\vert $ for all $m\in \mathcal{M}$ and all $n\in \mathcal{N}$, and

\item there exists $\hat{x}$ such that (i) $\hat{x}_{n}$ satisfies Condition
2 in Definition \ref{def:gamma_eq} and, (ii) $\hat{x}_{0}^{m}$ satisfies
Condition 3 in Definition \ref{def:gamma_eq} and, for all $m\in \mathcal{M}$%
, and all $\gamma ^{m}\in \Gamma ^{m}$,
\begin{gather}
\hat{x}_{0}^{m}\left( \gamma ^{m}\right) \in  \label{def:regular_1} \\
\arg \max_{c_{0}^{m}\in C_{0}^{m}}\left\{
\begin{array}{c}
G^{m}\left[ s\left( \hat{\sigma},\gamma ^{m},\tilde{\gamma}%
^{-m},c_{0}^{m},x_{-0}^{m},x^{-m}\right) \right] - \\
\sum_{n\in \mathcal{N}}\gamma _{n}^{m}(c_{0}^{m},\hat{x}_{-0}^{m}(\gamma
^{m},\tilde{\gamma}^{-m}))\left[ s_{n}\left( \hat{\sigma}_{n},\gamma ^{m},%
\hat{\gamma}^{-m},c_{0}^{m},\hat{x}_{-0}^{m},\hat{x}^{-m}\right) \right]%
\end{array}%
\right\} ,  \notag
\end{gather}%
where $\hat{\sigma}$ satisfies
\begin{gather}
\hat{\sigma}\left( \gamma ^{m},\tilde{\gamma}^{-m},c,t\right) =\left\{ \hat{%
\sigma}\left( \left( \gamma ^{m},\tilde{\gamma}^{-m}\right) ,c_{n},\left(
t^{m},\tilde{t}^{-m,m}\right) \right) \right\} _{n\in \mathcal{N}}
\label{def:regular_2} \\
\in \min_{s\in \hat{S}\left( t^{m},\tilde{t}^{-m,m}\right) }\left\{
G^{m}(s)-\sum_{n\in \mathcal{N}}t_{n}^{m}(s_{n})\right\}  \notag
\end{gather}%
for all $m\in \mathcal{M}$, all $n\in \mathcal{N}$, and all $\left( \gamma
^{m},c_{n},t\right) \in \Gamma ^{m}\times C_{n}\times T.$
\end{enumerate}
\end{definition}

The first condition in Definition \ref{def:regular} requires that the
cardinality of $C_{n}^{m}$ be weakly greater than the cardinality of $%
\overline{\mathcal{M}}$, which is $M+1$. The second condition is regarding
players' equilibrium communication strategies. First, $\hat{\sigma}$ is a
profile of agents' optimal action strategies. Note that $\hat{S}\left( t^{m},%
\tilde{t}^{-m,m}\right) $ is the set of all profiles of agents' optimal
actions when $\left( t^{m},\tilde{t}^{-m,m}\right) $ is a profile of
transfer schedules. (\ref{def:regular_2}) implies that if principals other
than $m$ offers $\tilde{\gamma}^{-m}$ so that their transfer schedules
become $\tilde{t}^{-m,m}$, agents chooses a profile of optimal actions that
generates the lowest payoff for principal $m$ among all profiles of optimal
actions in $\hat{S}\left( t^{m},\tilde{t}^{-m,m}\right) .$ The second
condition in Definition \ref{def:regular} requires the existence of agents'
communication strategies that satisfies Condition 2 in Definition \ref%
{def:gamma_eq} and the existence of each principal $m$'s communication
strategy that satisfies (\ref{def:regular_1}). This essentially ensures the
existence of equilibrium (pure) communication strategies upon any principal $%
m$'s deviation to any mechanism that triggers agents' punishments given
non-deviators' punishment transfer schedules $\tilde{t}^{-m,m}$.\footnote{%
Note that we restrict Condition 2 in Definition \ref{def:regular} to pure
communication strategies. However, it is not necessarry. We can apply it to
mixed communication strategies.}

\begin{theorem}
\label{thm:regular}If a game relative to $\Gamma $ is regular, $Z^{\Gamma
}=Z^{\ast }$ for any given $\Gamma $.
\end{theorem}

\begin{proof}
We first show that $Z^{\Gamma }\subset Z^{\ast }$. Fix any $\Gamma $
equilibrium $\left( \hat{\gamma},\hat{x},\hat{\sigma}\right) \in \mathcal{E}%
^{\Gamma }$. Note that a profile of agents' equilibrium actions $\left[ \hat{%
\sigma}_{n}\left( \hat{\gamma},\hat{x}_{n}(\hat{\gamma}),\hat{\gamma}(\hat{x}%
(\hat{\gamma}))\right) \right] _{n\in \mathcal{N}}$ is derived from their
payoff maximization problems given equilibrium transfer schedules $\hat{%
\gamma}(\hat{x}(\hat{\gamma})).$ (See Condition (i) in Definition \ref%
{def:gamma_eq}.) Therefore, $\left[ \hat{\sigma}_{n}\left( \hat{\gamma},\hat{%
x}_{n}(\hat{\gamma}),\hat{\gamma}(\hat{x}(\hat{\gamma}))\right) \right]
_{n\in \mathcal{N}}\in \hat{S}\left( \hat{\gamma}(\hat{x}(\hat{\gamma}%
))\right) ,$ which implies that qualifier (i) in $Z^{\ast }$ is satisfied.
Given $\hat{\gamma}^{m},$ suppose that principal $m$ unilaterally deviates
to a mechanism $\gamma ^{m}$ such that for all $c_{0}^{m}\in C_{0}^{m}$ and
all $c_{-0}^{m},\acute{c}_{-0}^{m}\in C_{-0}^{m}=\Pi _{n\in \mathcal{N}%
}C_{n}^{m}$%
\begin{eqnarray}
\gamma ^{m}\left( c_{0}^{m},c_{-0}^{m}\right) &=&\gamma ^{m}\left( c_{0}^{m},%
\acute{c}_{-0}^{m}\right)  \label{thm2_1} \\
\gamma ^{m}\left( C_{0}^{m},c_{-0}^{m}\right) &=&T^{m},  \label{thm2_2}
\end{eqnarray}%
where $\gamma ^{m}\left( C_{0}^{m},c_{-0}^{m}\right) \equiv \left\{ \gamma
^{m}\left( c_{0}^{m},c_{-0}^{m}\right) \in T^{m}:c_{0}^{m}\in
C_{0}^{m}\right\} .$ That is, agents' messages are irrelevant in determining
a profile of transfer schedules and principal $m$ can choose any profile of
transfer schedules he wants. When principal $m$ optimally chooses his
message, i.e., $\hat{x}_{0}^{m}\left( \gamma ^{m}\right) ,$ his continuation
equilibrium payoff upon deviation to $\gamma ^{m}$ satisfies the following
relations:%
\begin{gather}
G^{m}\left[ s\left( \hat{\sigma},\gamma ^{m},\hat{\gamma}^{-m},\hat{x}%
\right) \right] -\sum_{n\in \mathcal{N}}\gamma _{n}^{m}(\hat{x}^{m}(\gamma
^{m},\hat{\gamma}^{-m}))\left[ s_{n}\left( \hat{\sigma}_{n},\gamma ^{m},\hat{%
\gamma}^{-m},\hat{x}\right) \right] =  \label{thm2_3} \\
\max_{c_{0}^{m}\in C_{0}^{m}}\left\{
\begin{array}{c}
G^{m}\left[ s\left( \hat{\sigma},\gamma ^{m},\hat{\gamma}%
^{-m},c_{0}^{m},x_{-0}^{m},x^{-m}\right) \right] - \\
\sum_{n\in \mathcal{N}}\gamma _{n}^{m}(c_{0}^{m},\hat{x}_{-0}^{m}(\gamma
^{m},\hat{\gamma}^{-m}))\left[ s_{n}\left( \hat{\sigma}_{n},\gamma ^{m},\hat{%
\gamma}^{-m},c_{0}^{m},\hat{x}_{-0}^{m},\hat{x}^{-m}\right) \right]%
\end{array}%
\right\} \geq  \notag \\
\max_{t^{m}\in T^{m}}\left\{ \min_{s(t^{-m},t^{m})\in \hat{S}%
(t^{-m},t^{m})}G^{m}(s(t^{-m},t^{m}))-\sum_{n\in \mathcal{N}%
}t_{n}^{m}(s_{n}(t^{-m},t^{m}))\right\}  \notag
\end{gather}%
Note that the expression in the first line in (\ref{thm2_3}) is principal $m$%
's continuation equilibrium payoff upon deviation to $\gamma ^{m}.$ The
equality holds because of Condition 3 in Definition \ref{def:gamma_eq}. For
the inequality, first of all, note that given his deviation to $\gamma ^{m}$
satisfying (\ref{thm2_1}) and (\ref{thm2_2}), principal $m$ can induce any
profile of transfer schedules from $T^{m}$ with his message alone. Second,
given $\left( \gamma ^{m},\hat{\gamma}^{-m}\right) ,$ a profile of agents'
action choices $s(\hat{\sigma},\gamma ^{m},\hat{\gamma}^{-m},\hat{x})$ is in
$\hat{S}(\gamma ^{m}(\hat{x}^{m}(\gamma ^{m},\hat{\gamma}^{-m})),\hat{\gamma}%
^{-m}(\hat{x}^{-m}(\gamma ^{m},\hat{\gamma}^{-m})))$ because of Condition 1
in Definition \ref{def:gamma_eq} but not necessarily one that minimizes
principal $m$'s payoff. These two properties imply the inequality.

On the other hand, Condition 4 in Definition \ref{def:gamma_eq} implies that
\begin{gather}
G^{m}\left[ s\left( \hat{\sigma},\hat{\gamma},\hat{x}\right) \right]
-\sum_{n\in \mathcal{N}}\gamma _{n}^{m}(\hat{x}^{m}(\hat{\gamma}))\left[
s_{n}\left( \hat{\sigma}_{n},\hat{\gamma},\hat{x}\right) \right] \geq
\label{thm2_4} \\
G^{m}\left[ s\left( \hat{\sigma},\gamma ^{m},\hat{\gamma}^{-m},\hat{x}%
\right) \right] -\sum_{n\in \mathcal{N}}\gamma _{n}^{m}(\hat{x}^{m}(\gamma
^{m},\hat{\gamma}^{-m}))\left[ s_{n}\left( \hat{\sigma}_{n},\gamma ^{m},\hat{%
\gamma}^{-m},\hat{x}\right) \right] .  \notag
\end{gather}%
Combining (\ref{thm2_3}) and (\ref{thm2_4}), we conclude that the
equilibrium allocation satisfies qualifier (ii) in $Z^{\ast }$. Because both
qualifiers are satisfied, $z^{\left( \hat{\gamma},\hat{x},\hat{\sigma}%
\right) }\in Z^{\ast }$.

We now show that $Z^{\ast }\subset Z^{\Gamma }$. We pick any allocation $(%
\hat{s},\hat{d})\in Z^{\ast }$. Then, there exists $\hat{t}\in T\ $s.t.$\
\hat{s}\in \hat{S}(\hat{t})\ $and$\ \hat{d}_{n}^{m}=\hat{t}_{n}^{m}(\hat{s}%
_{n})\ $for all$\ (n,m)\in \mathcal{N}\times \mathcal{M}$ and each principal
$m$'s payoff, $G^{m}(\hat{s})-\sum_{n\in \mathcal{N}}\hat{d}_{n}^{m}$,
associated with $(\hat{s},\hat{d})$ is no less than $\underline{V}^{m}$ for
all $m\in \mathcal{M}$. Since the game is regular, $\left\vert
C_{n}^{m}\right\vert \geq \left\vert \overline{\mathcal{M}}\right\vert $.
Therefore, we can pick an arbitrary injective function $\phi _{n}^{m}:%
\overline{\mathcal{M}}\rightarrow C_{n}^{m}.$ Then, for any $k_{n}^{m}\in
\overline{\mathcal{M}}$, $\phi _{n}^{m}$ uniquely identifies a message $\phi
_{n}^{m}\left( k_{n}^{m}\right) $ in $C_{n}^{m}$. We construct each
principal $m$'s mechanism $\hat{\gamma}^{m}=\left\{ \hat{\gamma}%
_{n}^{m}\right\} _{n\in \mathcal{N}}$ as follows%
\begin{multline*}
\hat{\gamma}_{n}^{m}\left( c_{0}^{m},c_{1}^{m},\ldots ,c_{N}^{m}\right) = \\
\left\{
\begin{array}{cc}
\tilde{t}_{n}^{m,k} & \text{if }\exists k\neq m\text{ such that }\left\vert
\left\{ c_{n}^{m}\in C_{n}^{m}:c_{n}^{m}=\phi _{n}^{m}\left(
k_{n}^{m}\right) \text{, }k_{n}^{m}=k,\text{ }n\in \mathcal{N}\right\}
\right\vert >\frac{N}{2}, \\
&  \\
t_{n}^{\circ } &
\begin{array}{c}
\text{if }\exists k,k^{\prime }\neq m\text{ such that (i) }k\neq k^{\prime }%
\text{ and} \\
\text{(ii) }\left\vert \left\{ c_{n}^{m}\in C_{n}^{m}:c_{n}^{m}=\phi
_{n}^{m}\left( k_{n}^{m}\right) \text{, }k_{n}^{m}=k,\text{ }n\in \mathcal{N}%
\right\} \right\vert = \\
\left\vert \left\{ c_{n}^{m}\in C_{n}^{m}:c_{n}^{m}=\phi _{n}^{m}\left(
k_{n}^{m}\right) \text{, }k_{n}^{m}=k^{\prime },\text{ }n\in \mathcal{N}%
\right\} \right\vert =\frac{N}{2}%
\end{array}%
, \\
&  \\
\hat{t}_{n}^{m} & \text{otherwise.}%
\end{array}%
\right.
\end{multline*}%
Note that principal $m$'s message $c_{0}^{m}$ to himself is not relevant in
determining a profile of transfer schedules. Therefore, his message behavior
is not strategic, so one can fix some arbitrary message $\hat{x}%
_{n}^{m}\left( \hat{\gamma}^{m}\right) \in C_{n}^{m}.$ The mechanism $\hat{%
\gamma}^{m}$ resembles a DRM $\hat{\lambda}^{m}$ defined in (\ref{thm1_1B}).
The difference is that we relabel agents' messages with ones in the bigger
message set $C^{m}$ with no role of principal $m$'s messages in determining
a profile of transfer schedules. For all $n\in \mathcal{N}$ and all $k\in
\mathcal{M\diagdown }\left\{ m\right\} ,$ let agents report the message $%
\hat{x}_{n}^{m}\left( \gamma ^{k},\hat{\gamma}^{-k}\right) $ to all $m\neq k$
such that
\begin{equation}
\hat{x}_{n}^{m}\left( \gamma ^{k},\hat{\gamma}^{-k}\right) =\left\{
\begin{array}{cc}
\phi _{n}^{m}\left( k\right) & \text{if }\gamma ^{k}\neq \hat{\gamma}^{k};
\\
\phi _{n}^{m}\left( 0\right) & \text{if }\gamma ^{k}=\hat{\gamma}^{k}.%
\end{array}%
\right.  \label{thm2_5}
\end{equation}%
The strategy of communicating with non-deviating principal $m$ in (\ref%
{thm2_5}) is supported as a truthful equilibrium communication because of
the structure of $\hat{\gamma}^{m}$. If no principal deviates, the profile
of these communication strategies induce $\hat{t}=\left\{ \hat{t}%
^{m}\right\} _{m\in \mathcal{M}}$. We choose a profile of agents' action
strategies on the equilibrium path such that
\begin{equation}
\hat{\sigma}\left( \hat{\gamma},\hat{x},\hat{\gamma}\left( \hat{x}\right)
\right) =\hat{s}\text{.}  \label{thm2_6}
\end{equation}%
Because $\hat{s}\in \hat{S}(\hat{t}),$ $\hat{\sigma}\left( \hat{\gamma},\hat{%
x},\hat{\gamma}\left( \hat{x}\right) \right) $ is a profile of agents'
optimal actions given their truthful reports $\phi _{n}^{m}\left( 0\right) $
for all $m\in \mathcal{M}$ and all $n\in \mathcal{N}$. Because $\hat{\sigma}%
\left( \hat{\gamma},\hat{x},\hat{\gamma}\left( \hat{x}\right) \right) =\hat{s%
}$ and agents report truthfully, we also have%
\begin{equation}
\hat{t}_{n}^{m}(\hat{\sigma}_{n}\left( \hat{\gamma},\hat{x}_{n}\left( \hat{%
\sigma}\left( \hat{\gamma},\hat{x},\hat{\gamma}\left( \hat{x}\right) \right)
\right) ,\hat{\gamma}\left( \hat{x}\right) \right) )=\hat{d}_{n}^{m}
\label{thm2_7}
\end{equation}%
for all $m\in \mathcal{M}$ and all $n\in \mathcal{N}$.

Therefore, $(\hat{s},\hat{d})\in Z^{\Gamma }$ if we can show that each
principal $m$ does not gain in a continuation equilibrium upon his deviation
to any other mechanism in $\Gamma ^{m}$. Given agents' truthful reporting to
non-deviating principals, non-deviators' transfer schedules become $\tilde{t}%
^{-m,m}$ upon principal $m$'s deviation to $\acute{\gamma}^{m}\neq \hat{%
\gamma}^{m}.$ Assume that agents choose actions that minimizes $%
G^{m}(s)-\sum_{n\in \mathcal{N}}\acute{t}_{n}^{m}(s_{n})$ among all optimal
actions in $\hat{S}(\tilde{t}^{-m,m},\acute{t}^{m})$ given any profile of
transfer schedules $\acute{t}^{m}$ that are eventually assigned by deviating
principal $m$ in a continuation equilibrium upon principal $m$'s deviation
to $\acute{\gamma}^{m}\neq \hat{\gamma}^{m}$. The existence of a
continuation equilibrium upon principal $m$'s deviation to $\acute{\gamma}%
^{m}\neq \hat{\gamma}^{m}$ is ensured because of the second requirement for
a \textquotedblleft regular\textquotedblright\ game. Then, agents' action
optimal choice implies that principal $m$'s payoff is $\min_{s\in \hat{S}(%
\tilde{t}^{-m,m},t^{-m})}G^{m}(s)-\sum_{n\in \mathcal{N}}\acute{t}%
_{n}^{m}(s_{n})$ and it satisfies
\begin{eqnarray}
&&\min_{s\in \hat{S}(\tilde{t}^{-m,m},\acute{t}^{m})}\left[
G^{m}(s)-\sum_{n\in \mathcal{N}}\acute{t}_{n}^{m}(s_{n})\right]
\label{thm2_8} \\
&\leq &\min_{t^{-m}\in T^{-m}}\max_{t^{m}\in T^{m}}\left\{
\min_{s(t^{-m},t^{m})\in \hat{S}(t^{-m},t^{m})}G^{m}(s(t^{-m},t^{m}))-%
\sum_{n\in \mathcal{N}}t_{n}^{m}(s_{n}(t^{-m},t^{m}))\right\} =\underline{V}%
^{m}  \notag
\end{eqnarray}%
where the inequality holds because of the definition of $\tilde{t}^{-m,m}$
in (\ref{thm1_1A}) and $\acute{t}^{m}\in T^{m}.$ On the other hand,
principal $m$'s payoff on the equilibrium path with $\hat{\gamma}^{m}$ is
the expression on the left-hand side of the inequality relation below
because of (\ref{thm2_5}) - (\ref{thm2_7}):
\begin{equation}
G^{m}(\hat{s})-\sum_{n\in \mathcal{N}}\hat{d}_{n}^{m}\geq \underline{V}^{m},
\label{thm2_9}
\end{equation}%
where the inequality relation holds because on the equilibrium path because $%
(\hat{s},\hat{d})\in Z^{\ast }$. Finally, combining (\ref{thm2_8}) and (\ref%
{thm2_9}) yields
\begin{equation*}
G^{m}(\hat{s})-\sum_{n\in \mathcal{N}}\hat{d}_{n}^{m}\geq \min_{s\in \hat{S}(%
\tilde{t}^{-m,m},\acute{t}^{m})}\left[ G^{m}(s)-\sum_{n\in \mathcal{N}}%
\acute{t}_{n}^{m}(s_{n})\right] .
\end{equation*}%
Therefore, principal $m$ does not gain upon deviation to any $\acute{\gamma}%
^{m}\neq \hat{\gamma}^{m}$. Therefore, $(\hat{s},\hat{d})\in Z^{\Gamma }$.
\end{proof}

For Theorem \ref{thm:regular}, it is the key to understand the role of a
principal's message to himself. Given a mechanism $\gamma ^{m}$, principal $%
m $ can make his offer of transfer schedules conditional on his message. One
extreme is that his offer of transfer schedules depend only on his message
and he can offer any profile of transfer schedules in $T^{m}$ by sending
some message to himself. In a continuation equilibrium upon deviation to
such a mechanism, he will choose a profile of transfer schedules that
maximizes his payoff given his belief on agents' optimal action choices
conditional on transfer schedules emerging from non-deviators' mechanisms.
Therefore, principal $m$'s equilibrium payoff cannot be lower than the
minmax value of his payoff over transfer schedules calculated on the basis
of agents' action choices such that given each profile of transfer
schedules, they choose optimal actions that minimize principal $m$'s payoff
among all possible optimal actions. Therefore, principal $m$'s equilibrium
payoff cannot be lower than $\underline{V}^{m}$, that is, an equilibrium
allocation must be in $Z^{\ast }$

Suppose that a game relative to $\Gamma $ is regular. Then, for any
allocation in $Z^{\ast },$ we can construct an equilibrium mechanism for
each principal, which resembles a DRM that ensures that any principal $m$
cannot gain upon his deviation to any mechanism in $\Gamma ^{m}$. This is
how Theorem \ref{thm:regular} is established.

\subsection{Appendix F}

The following corollary shows that any equilibrium allocation in a GPTA is
robust in the sense that it does not disappear even if we allow principals
to offer DRMs.

\begin{corollary}
\label{coroll_GPTA}$Z^{G}\subset Z^{\ast }$
\end{corollary}

\begin{proof}
See Appendix \ref{sec:pf:corollary}.
\end{proof}

\bigskip

Recall that $Z^{G}$ is the set of equilibrium allocation in a GPTA. It is
clear that $Z^{G}\subset Z^{\ast }$ because principals are individually
rational subject to agents' action choices given transfer schedules in an
equilibrium of a GPTA. Therefore, the equilibrium allocation in a GPTA is
robust to the possibility that a principal may offer a DRM. In fact, the
equilibrium allocation in a GPTA survives even when a principal deviates to
more complex mechanisms that DRMs because the set of equilibrium allocations
with any arbitrarily complex mechanisms is the same as $Z^{\ast }$ (See
Theorem \ref{thm:regular} in Appendix \ref{sec:CMGPTA_Gamma}).

While, there are generally new allocations that are supported by truthful $%
\Lambda $ equilibria. Furthermore, it is not necessary for the equilibrium
transfer schedules in a $\Lambda $ equilibrium to be weakly truthful. On the
other hand, equilibrium transfer schedules in a GPTA must be weakly truthful
if each agent has only two actions and cares solely about monetary payoff

\subsection{Appendix G}

Table \ref{tab:sumstats} provides a summary of demographic information by
session type.

\begin{table}[h]
\centering
\begin{tabular}{lll}
& \textbf{Computer Agents} & \multicolumn{1}{c}{\textbf{Human Agents}} \\
\textbf{Median Age} & \multicolumn{1}{c}{22} & \multicolumn{1}{c}{24.3} \\
&  &  \\
\textbf{Gender} &  &  \\
\multicolumn{1}{l}{Male} & \multicolumn{1}{c}{47.9\%} & \multicolumn{1}{c}{
33.8\%} \\
\multicolumn{1}{l}{Female} & \multicolumn{1}{c}{50.7\%} & \multicolumn{1}{c}{
61.9\%} \\
\multicolumn{1}{l}{Other} & \multicolumn{1}{c}{1.4\%} & \multicolumn{1}{c}{
4.3\%} \\
&  &  \\
\textbf{Student} & \multicolumn{1}{c}{97.2\%} & \multicolumn{1}{c}{85.7\%}
\\
&  &  \\
\textbf{Field of Study} &  &  \\
\multicolumn{1}{l}{Business, Social Sciences and Humanities} &
\multicolumn{1}{c}{22.9\%} & \multicolumn{1}{c}{24.2\%} \\
\multicolumn{1}{l}{STEM} & \multicolumn{1}{c}{77.1\%} & \multicolumn{1}{c}{
75.8\%} \\
&  &  \\
\textbf{Median Understanding (1-7)} & \multicolumn{1}{c}{5} &
\multicolumn{1}{c}{5}%
\end{tabular}%
\par
\vspace{\baselineskip} {\footnotesize \raggedright \textit{Note: Gender,
Student, and Field of Study are given as percentages of respondents. Fields
of Study are grouped into STEM (Engineering, Sciences or Health Sciencesin
our data) or Business, Social Sciences and Humanities. Median Understanding
is the median reported level of understanding of the instructions and
experiment from a scale of 1 to 7, with 7 representing full and complete
understanding.} }
\caption{Summary Statistics by Treatment}
\label{tab:sumstats}
\end{table}

\newpage

\subsection{Appendix H}

\begin{figure}[]
\centering
\includegraphics[width=.8\textwidth]{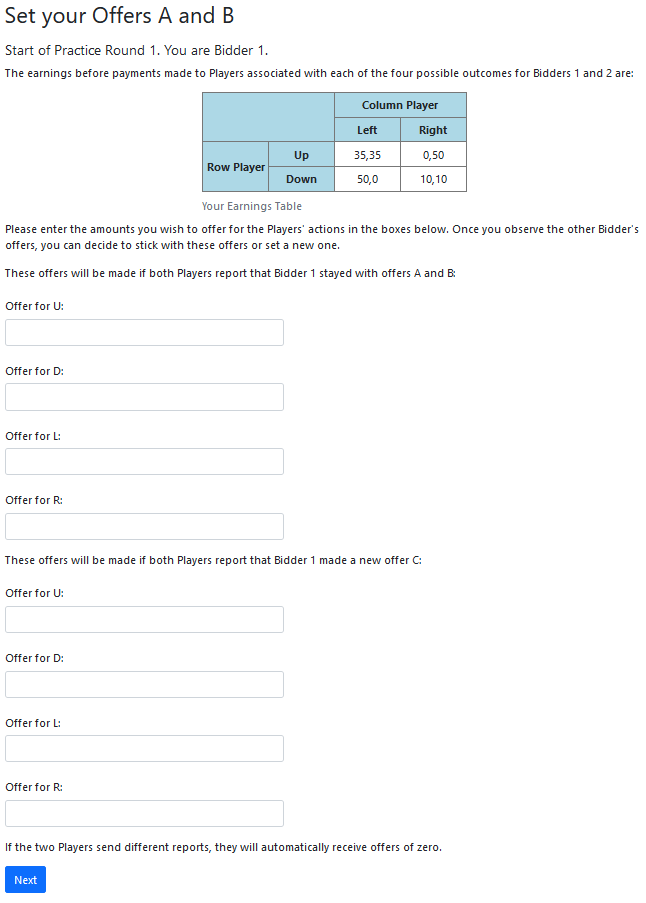}
\caption{Set Offers AB Screen}
\label{fig:offersAB}
\end{figure}

\begin{figure}[]
\centering
\includegraphics[width=\textwidth]{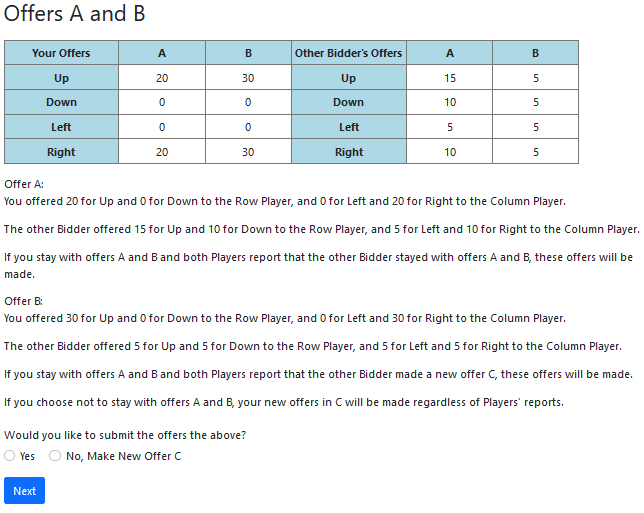}
\caption{Deviation Choice Screen}
\label{fig:devchoiceshot}
\end{figure}

\begin{figure}[]
\centering
\includegraphics[width=\textwidth]{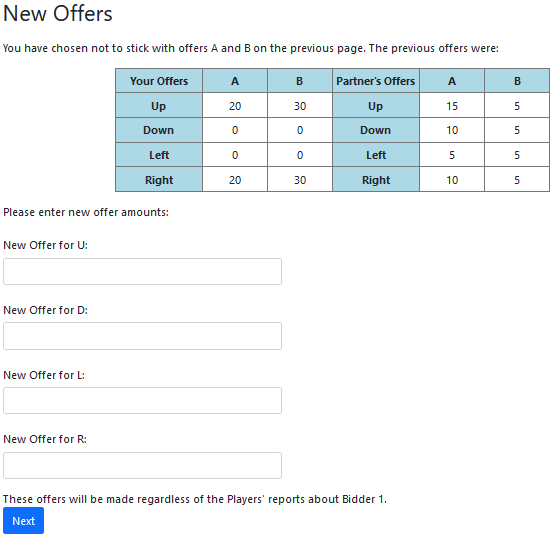}
\caption{Set Offers C Screen}
\label{fig:offersC}
\end{figure}

\begin{figure}[]
\centering
\includegraphics[width=\textwidth]{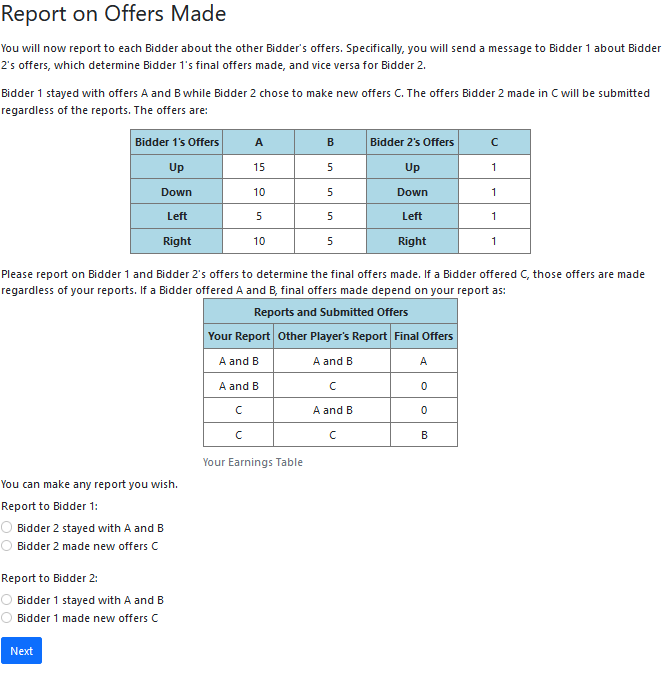}
\caption{Agent's Report Screen}
\label{fig:reports}
\end{figure}

\begin{figure}[]
\centering
\includegraphics[width=\textwidth]{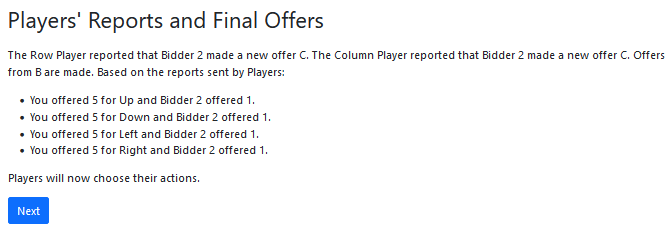}
\caption{Submitted Offers Information Screen for Bidders}
\label{fig:submitted}
\end{figure}

\begin{figure}[]
\centering
\includegraphics[width=.8\textwidth]{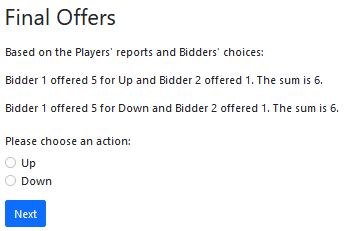}
\caption{(Row) Agent's Action Screen}
\label{fig:action}
\end{figure}

\newpage

\subsection{Appendix H}

\begin{figure}[tbp]
\centering
\includegraphics[width=1\textwidth]{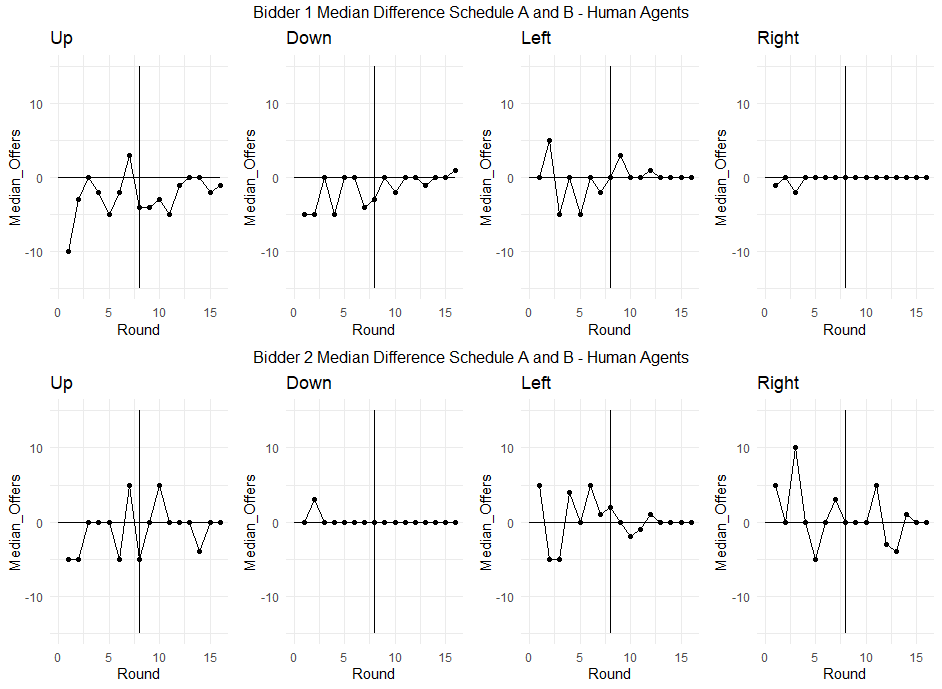}
\caption{Median Differences in offers in Schedule B minus Schedule A }
\label{fig:T2_med_diffs}
\end{figure}

Offers between Schedule profiles A and B are not very different. Figure \ref%
{fig:T1_med_diffs} shows the difference in offers at the median between
Schedule B and Schedule A with Computer agents and \ref{fig:T2_med_diffs}
with Human agents. A positive amount indicates that the offer for a
particular action is greater in Schedule B, the deviator-punishing transfer
schedule, than in Schedule A.

\end{document}